\documentstyle[prb,eqsecnum,aps,epsf]{revtex}
\tighten

\begin{document}
\draft \title{Magnetoresistance of Granular Superconducting Metals in
a Strong Magnetic Field} \author{I.S. Beloborodov$^{(1)}$,
K.B. Efetov$^{(1,2)}$, A.I. Larkin$^{(3,2,1)}$}
\address{$^{(1)}$Theoretische Physik III,\\ Ruhr-Universit\"{a}t
Bochum, 44780 Bochum, Germany\\ $^{(2)}$L.D. Landau Institute for
Theoretical Physics, 117940 Moscow, Russia \\ $^{(3)}$Theoretical
Physics Institute, University of\\ Minnesota, Minneapolis, MN 55455,
USA} \date{\today} \maketitle

\begin{abstract}
The magnetoresistance of a granular superconductor in a strong
magnetic field is considered. It is assumed that this field destroys
the superconducting gap in each grain, such that all interesting
effects considered in the paper are due to superconducting
fluctuations. The conductance of the system is assumed to be large,
which allows us to neglect all localization effects as well as the
Coulomb interaction. It is shown that at low temperatures the
superconducting fluctuations reduce the one-particle density of states
but do not contribute to transport. As a result, the resistivity of
the normal state exceeds the classical resistivity approaching the
latter only in the limit of extremely strong magnetic fields, and this
leads to a negative magnetoresistance. We present detailed
calculations of physical quantities relevant for describing the effect
and make a comparison with existing experiments.
\end{abstract}


%\begin{multicols}{2}

\section{Introduction}

In a recent experiment \cite{Gerber97}, transport properties a system
of $Al$ superconducting grains in a strong magnetic field were
studied. The samples were quite homogeneous with a typical diameter of
the grains $120\pm 20\AA $ and the grains formed a $3$-dimensional
array. As usual, sufficiently strong magnetic fields destroyed the
superconductivity in the samples and a finite resistivity could be
seen above a critical magnetic field. The applied magnetic fields
reached $17T$, which was more than sufficient to destroy also the
superconducting gap in each grain.

The dependence of the resistivity on the magnetic field observed in
Ref.~ \cite{Gerber97} was not simple. Although at extremely strong
fields the resistivity was almost independent of the field, it {\it
increased }when decreasing the magnetic field. Only at sufficiently
weak magnetic fields the resistivity started to decrease and finally
the samples displayed superconducting properties. A similar behavior
had been reported in a number of publications
\cite{Rutgers,Gantmakher96}.

A negative magnetoresistance due to weak localization effects is not
unusual in disordered metals \cite{lee}. However, the
magnetoresistance of the granulated materials studied in
Ref. \cite{Gerber97} is quite noticeable in magnetic fields exceeding
$10T$ and is considerably larger than values estimated for the weak
localization.

\ The weak localization effects become very important if the system is
near the Anderson metal-insulator transition and one can expect there
a complicated dependence on the magnetic field. One can, for example,
argue \cite{Rutgers} that decreasing the magnetic field drives the
system to the metal-insulator transition, which gives a large negative
magnetoresistance. If the superconducting transition occurs
earlier one gets a maximum of the resistivity near the transition
characteristic for the experiments
\cite{Gerber97,Rutgers,Gantmakher96}.

Unfortunately, investigation of this possibility is not simple. The
metal-insulator transition occurs at values of the macroscopic
conductance $J $ of the order of unity. At such values calculations
are very difficult. The problem becomes even more complicated due to
the Coulomb interaction. It is well known \cite{efetov80}, that, at
small values of $J$, the system must be an insulator even if the
superconducting gap is finite in a single grain. A microscopic
consideration of all these effects and a confirmation of the existence
of the negative magnetoresistance near the superconducting point for
$J\sim 1$ is at present hardly possible and we do not try to treat the
problem here.

Instead, we consider below the region of large conductances $J\gg 1$,
where the system without interactions would be a good metal. This
region corresponds to large tunneling amplitudes between the
grains. All effects of the weak localization and the charging effects
have to be small for $J\gg 1$, which would imply that the
resistivity could not considerably depend on the magnetic field.

Nevertheless, we find that the magnetoresistance of a good granulated
metal ($J\gg 1$) in a strong magnetic field and at low temperature
{\it must be} {\it negative.} In our model, the superconducting gap in
each granule is assumed to be suppressed by the strong magnetic
field. All the interesting behavior considered below originates from
the superconducting fluctuations that lead to a suppression of the
density of states (DOS) but do not help to carry an electric
current. The main results have been already published \cite {Igor} and
now we want to present details of calculations and clarify some
additional questions. We consider, for example, influence of the
Zeeman splitting on the resistivity, find the critical field $H_{C2}$
of the transition into the superconducting state, calculate the
diamagnetic susceptibility and estimate weak localization corrections.

Theory of superconducting fluctuations near the transition into the
superconducting state has been developed long ago \cite
{Aslamazov68,Maki68,Abrahams} (for a review see
Ref.~\onlinecite{Varlamov}).  Above the transition temperature
$T_{c}$, non-equilibrium Cooper pairs are formed and a new channel of
charge transfer opens (Aslamazov-Larkin
contribution)\cite{Aslamazov68}. Another fluctuation contribution
comes from a coherent scattering of the electrons forming a Cooper
pair on impurities (Maki-Thompson contribution)\cite{Maki68}. Both the
fluctuation corrections increase the conductivity and, when a magnetic
field is applied, lead to a positive magnetoresistance. Formation of
the non-equilibrium Cooper pairs results also in a fluctuational gap
in the one-electron spectrum \cite {Abrahams} but in conventional (non
granular) superconductors the first two mechanisms are more important
near $T_{c}$ and the conductivity increases when approaching the
transition. The total conductivity for a bulk sample above the
transition temperature $T_{c}$ can be written in the following form
\begin{equation}
\sigma =\sigma _{Drude}+\delta \sigma ,  \label{Drud}
\end{equation}
where $\sigma _{Drude}=(e^{2}\tau n)/m$ is the conductivity of a
normal metal without electron-electron interaction, $\tau $ is the
elastic mean free time, $m$ and $n$ are the effective mass and the
density of electrons, respectively. In Eq. (\ref{Drud}), $\delta
\sigma $ is a correction to the conductivity due to the fluctuations
of the virtual cooper pairs
\begin{equation}
\delta \sigma =\delta \sigma _{DOS}+\delta \sigma _{AL}+\delta \sigma _{MT}
\end{equation}
where $\delta \sigma _{DOS}$ is the correction to the conductivity due
to the reduction of the DOS and $\delta \sigma _{AL}$ and $\delta
\sigma _{MT}$ stand for the Aslamazov-Larkin (AL) and Maki-Thompson
(MT) contributions the conductivity. Close to the critical temperature
$T_{c}$ the AL correction is more important than both the MT and DOS
corrections types and its contribution can be written as follows
\cite{Aslamazov68}
\begin{equation}
\frac{\delta \sigma _{AL}}{\sigma _{Drude}}=\lambda \left(
\frac{T_{c}}{T-T_{c}}\right) ^{\beta } \label{AL1}
\end{equation}
where $\lambda $ is a small dimensionless positive parameter $\lambda
\ll 1$ and $\beta =1/2$ for the three-dimensional case (3D), 1 for 2D
and 3/2 for quasi-1D. Eq. (\ref{AL1}) was derived using a perturbation
theory and therefore is valid provided the inequality $\delta \sigma
_{AL}/\sigma _{Drude}\ll 1$ is fulfilled.

Although typically the AL and MT corrections are larger than the DOS
contribution, a small decrease of the transverse conductivity is
possible in layered materials \cite{ioffe} in a temperature interval
not very close to the transition. It is relevant to emphasize that all
previous study of the fluctuations has been done near the critical
temperature $T_{c}$ in a zero or a weak magnetic field. In contrast,
we are mainly concentrated on study of the transport at very low
temperature in a strong magnetic field. To the best of our knowledge,
fluctuations in this region have not been considered so far.

A strong magnetic field destroys the superconducting gap in each
granule.  However, even at magnetic fields $H$ exceeding the critical
field $H_{c}$ virtual Cooper pairs can still be formed. It turns out,
and it will be shown below, the influence of these pairs on the
macroscopical transport is drastically different from that near
$T_{c}$. The existence of the virtual pairs leads to a reduction of
the DOS but, in the limit $T\rightarrow 0,$ these pairs cannot travel
from one granule to another. As a result, the conductivity $\sigma $
can be at $H>H_{c}$ considerably lower than conductivity $\sigma _{0}$
of the normal metal without an electron-electron interaction. It
approaches the value $\sigma _{0}$ only in the limit $H\gg H_{c},$
when all the superconducting fluctuations are completely suppressed by
the magnetic field.

The superconducting pairing inside the grains is destroyed by both the
orbital mechanism and the Zeeman spliting. The critical magnetic field
$H_{c}^{or}$ destroying the superconductivity in a single grain in
this case can be estimated as $H_{c}^{or}R\xi \approx \phi _{0}$,
where $\phi _{0}=hc/e $ is a flux quantum, $R$ is a radius of single
grain and $\xi =\sqrt{\xi _{0}l}$ is the superconducting coherence
length. The Zeeman critical magnetic field $H_{c}^{z}$ can be written
as $g\mu _{B}H_{c}^{z}=\Delta _{0}$, where $\Delta _{0}$ is a BCS gap
for the single grain at magnetic field $H=0$ and $g$ is a Lande
factor. We notice here that $H_{c}^{z}$ is independent of the size of
the grain whereas for $H_{c}^{or}$ the size of the grain is
important. The ratio of this two fields can be written in the form
\begin{equation}
H_{c}^{or}/H_{c}^{z}\approx R_{c}/R, \label{orbit}
\end{equation}
where $R_{c}=\xi (p_{0}l)^{-1}$. We can see from Eq. (\ref{orbit})
that for $R>R_{c}$ the orbital critical magnetic field is smaller
than the Zeeman critical magnetic field $H_{c}^{or}<H_{c}^{z}$ and the
suppression of superconductivity is due to the orbital mechanism. This
condition is well satisfied in grains with $R\sim 100\AA $ studied in
\cite{Gerber97}. This limit is opposite to the one considered recently
in Ref. \cite{Aleiner97} where the Zeeman splitting was assumed to be
the main mechanism of destruction of the Cooper pairs. However, the
latter mechanism of the destruction of the Cooper pairs can be easily
included into the scheme of our calculations.

The remainder of the paper is organized as follows. In Sec.~\ref{sec1}
we formulate the model and discuss the fluctuational contributions to
the total conductivity of the granulated
superconductors. Sec. ~\ref{sec2} contains the derivation of the
correction to the conductivity of granulated superconductors due to
single electron tunnelling (DOS contribution) at very low temperatures
$T\ll T_{c}$ and strong magnetic fields $H-H_{c}\ll H_{c}$.  In
Sec. ~\ref{sec3} and Sec. \ref{sec4} corrections to the conductivity
due to tunneling of virtual Cooper pairs are derived (Aslamazov-Larkin
and Maki-Thompson corrections). In Sec. \ref{sec6.5} we discuss the
influence of magnetic field on the phase of order parameter and
calculate the critical field of the transition into the
superconducting phase. The importance of the Zeeman splitting is
discussed in Sec. \ref{6.6}. The contribution of fluctuating Cooper
pairs to the diamagnetic susceptibility of granular system is derived
in Sec.\ref{sec6}. Sec.~\ref{sec5} includes the results for
conductivity at $T-T_{c}\ll T_{c}$ and $H\ll H_{c}$. A discussion of a
recent experiment in $Al$ grains and a comparison with the theory is
presented in Sec.~\ref{sec7} Our results are summarized in the
Conclusion.
\begin{figure}
\epsfysize =5cm
\centerline{\epsfbox{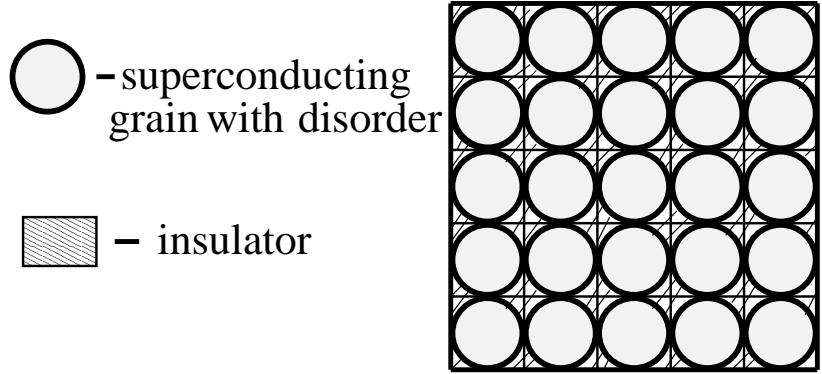}}
\caption{System of metallic grains}
\label{fig1}
\end{figure}

\section{Choice of the model}

\label{sec1}

We consider a 3D array of superconducting grains coupled to each
other, Fig.  \ref{fig1}. The grains are not perfect and there can be
impurities inside the grains as well as on the surface. We assume that
electrons can hop from grain to grain and can interact with phonons.

The Hamiltonian $\hat{H}$ of the system can be written as 
\begin{equation}
\hat{H}=\hat{H}_{0}+\hat{H}_{T},  \label{fulH}
\end{equation}
where $\hat{H}_{0}$ is a conventional Hamiltonian for a single grain
with an electron-phonon interaction in the presence of a strong
magnetic field
\begin{equation}
\hat{H}_{0}=\sum\limits_{i,k}E_{i,k}a_{i,k}^{\dag }a_{i,k}-|\lambda
|\sum\limits_{i,k,k^{\prime }}a_{i,k}^{\dag }a_{i,-k}^{\dag
}a_{i,-k^{\prime }}a_{i,k^{\prime }}+\hat{H}_{imp}, \label{hamilton}
\end{equation}
where $i$ stands for the numbers of the grains, $k\equiv ({\bf
k},\uparrow )$, $-k\equiv (-{\bf k},\downarrow )$; $\lambda $ is an
interaction constant, and $\hat{H}_{imp}$ describes elastic
interaction of the electrons with impurities. The interaction in
Eq. (\ref{hamilton}) contains diagonal matrix elements only. This form
of the interaction can be used provided the superconducting gap
$\Delta _{0}$ is not very large
\begin{equation}
\Delta _{0}\ll E_{c}  \label{thou}
\end{equation}
where $E_{c}$ is the Thouless energy of the single
granule. Eq. (\ref{thou}) is equivalent to the condition $R\ll \xi
_{0}$, where $R$ is the radius of the grain and $\xi _{0}$ is the
superconducting coherence length. In this limit, superconducting
fluctuations in a single grain are zero-dimensional.

The term $\hat{H}_{T}$ in Eq.(\ref{fulH}) describes tunneling from grain to
grain and has the form (see e.g. Ref.~\cite{Kulik}) 
\begin{equation}
\hat{H}_{T}=\sum\limits_{i,j,p,q}t_{ijpq}a_{ip}^{\dag }a_{jq}\exp
(i\frac{e}{c}{\bf A}{\bf d}_{ij})+h.c.  \label{tun}
\end{equation}
where ${\bf A}$ is the external vector potential; ${\bf d}_{ij}$ are
the vectors connecting centers of two neighboring grains $i$ and $j$
($\left| {\bf d}_{ij}\right| =2R$); $a_{ip}^{\dag }\left(
a_{ip}\right) $ are the creation (annihilation) operators for an
electron the grain $i$ and state $p$.

It is assumed that the system is macroscopically a good metal and this
corresponds to a sufficiently large tunneling energy $t$
\begin{equation}
t\gg \delta \label{a1}
\end{equation}
where $\delta =\left( \nu _{0}V\right) ^{-1}$ is the mean level
spacing in a single granule, $\nu _{0}=mp_{0}/2\pi ^{2}$ is the DOS
per one spin in the absence of interactions, $V$ is the volume of the
granule, and $\Delta _{0}$ is the Bardeen-Cooper-Schrieffer (BCS) gap
at $T=0$ in the absence of a magnetic field. Provided the inequality
(\ref{a1}) is fulfilled localization effects can be neglected
Ref.~\cite{Efetov}. Moreover, charging effects are also not important
in this limit because at such tunneling energies the Coulomb
interaction is well screened.

The tunneling current operator is 
\begin{equation}
\hat{{\bf j}}=-ie{\bf d}\sum\limits_{j,p,q}t_{ijpq}a_{ip}^{\dag
}a_{jq}\exp (i\frac{e}{c}{\bf A}{\bf d}_{ij})-h.c.
\end{equation}

Using standard formulae of the linear response theory we can write the
current ${\bf j}\left( t\right) $ in the form
\begin{equation}
{\bf j}(t)=i\int\limits_{-\infty }^{t}\langle \lbrack {\bf
\hat{\jmath}}(t), {\bf \hat{\jmath}}(t^{\prime })]\rangle {\bf
A}(t^{\prime })dt^{\prime }-ie^{2}{\bf d}^{2}\int\limits_{-\infty
}^{t}\langle \lbrack \hat{H}_{T}(t), \hat{H}_{T}(t^{\prime })]\rangle
dt^{\prime }{\bf A}(t), \label{j}
\end{equation}
where the angle brackets stand for averaging over both quantum states
and impurities in the grains. All operators in the right hand side of
Eq. (\ref {j}) are independent of the vector potential. In principle,
the grains can be clean and the electrons can scatter mainly on the
surface of the grains.  However, provided the shape of the grains
corresponds to a classically chaotic motion of the electrons, the
clean limit should be described in the $0D$ case by the same
formulae.

We carry out the calculation of the conductivity making expansion both
in fluctuation modes and in the tunneling term $H_{T}$. This implies
that the tunneling energy $t$ is not very large. Proper conditions
will be written later but now we mention only that the tunneling
energy $t$ will be everywhere much smaller than the energy $E_{c}$.

As in conventional bulk superconductors we can write corrections to
the classical conductivity as a sum of corrections to the DOS and of
Aslamazov-Larkin (AL) $\sigma _{AL}$ and Maki-Thompson (MT) $\sigma
_{MT}$ corrections. Diagrams describing these contributions are
represented in Fig.  2.
\begin{figure}
\epsfysize =4cm
\centerline{\epsfbox{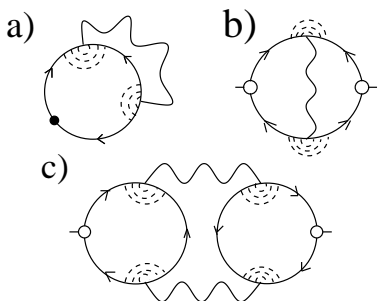}}
\caption{Diagram a) describes correction to DOS, diagrams b) and c)
describe corrections to conductivity due to superconducting
fluctuations. The wavy lines denote the propagator of the
fluctuations, the dashed lines stand for the impurity scattering.}
\end{figure}
The total conductivity $\sigma $ can be written as
\begin{equation}
\sigma =\sigma _{DOS}+\sigma _{AL}+\sigma _{MT}  \label{a2}
\end{equation}
where $\sigma _{DOS}$ is given by equation
\begin{equation}
\sigma _{DOS}=\sigma _{0}\left( 4T\right) ^{-1}\int\limits_{-\infty
}^{+\infty }[\nu (\varepsilon )/\nu _{0}]^{2}\cosh
^{-2}(\frac{\varepsilon }{2T})d\varepsilon , \label{a3}
\end{equation}
In Eq. (\ref{a3}), $\sigma _{0}=2\pi e^{2}R^{-1}\left( t/\delta
\right) ^{2}$ is the classical conductivity of the granular metal. It
can be rewritten in terms of the dimensionless conductance of the
system $J$ as
\begin{equation}
\sigma _{0}=\frac{8Je^{2}}{\pi R}  \label{a4}
\end{equation}
where the conductance $J$ equals 
\begin{equation}
J\equiv (\pi ^{2}/4)(t/\delta )^{2} \label{a40}
\end{equation}

The inequality (\ref{a1}) is equivalent to the condition
\begin{equation}
J\gg 1  \label{a400}
\end{equation}
The function $\nu \left( \varepsilon \right) $ in Eq. (\ref{a3}) is
the density of states. Without the electron-electron interaction, this
function is equal to the DOS of the ideal electron gas $\nu _{0}$,
which gives $\sigma _{DOS}=\sigma _{0}$. Taking into account the
electron-electron attraction we can write the contribution $\sigma
_{DOS}$ to the classical conductivity as
\begin{equation}
\sigma _{DOS}=\sigma _{0}+\delta \sigma _{DOS}(t,T,H)
\end{equation}
The correction $\delta \sigma _{DOS}\left( t,T,H\right) $ depends on
temperature $T$ and magnetic field $H$ and is represented in
Fig. 2a. As concerns the long range part of the Coulomb interaction
(charging effects), the condition $J\gg 1$ allows us to neglect it.

Using Eq. (\ref{a3}) the correction to the conductivity $\delta \sigma
_{DOS} $ at low temperatures can be written in terms of the correction
to the DOS at zero energy $\delta \nu \left( 0\right) $ as
\begin{equation}
\delta \sigma _{DOS}/\sigma _{0}=2(\delta \nu \left( 0\right) /\nu
_{0})
\label{connection}
\end{equation}
As we will see below, the main contribution to the conductivity due to
the superconducting fluctuations comes from the change of the DOS
$\delta \nu $.  Its calculation will be presented in detail in the
next Section.

\section{\protect\bigskip Suppression of the conductivity due to DOS
fluctuations}

\label{sec2}

In this Section we consider the correction to the conductivity of
granulated superconductors due to suppression of DOS. The main
correction $\delta \nu \left( \varepsilon \right) $ to the DOS of the
non-interacting electrons $\nu _{0}$ is described by the diagram in
the Fig. 2a, while the terms $\ \sigma _{MT}$ and $\sigma _{AL}$ are
given by Figs. 2b and 2c, respectively.  The calculation of the
diagrams can be performed for the Matsubara frequencies $\varepsilon
_{n}=\pi T(2n+1)$ using temperature Green functions. At the end one
should, as usual Ref.~\cite{AGD}, make the analytical continuation
$i\varepsilon _{n}\rightarrow \varepsilon $ . The magnetic field will
be considered in the quasi-classical approximation $l\ll L_{c}$, where
$l$ is a mean free path and $L_{c}$ is a cyclotron radius. In this
approximation, the magnetic field results in the appearance of
additional phases in Green functions
\begin{equation}
G(i\epsilon _{n},\vec{r},\vec{r^{\prime }})=G^{(0)}(i\epsilon
_{n},\vec{r}-\vec{r^{\prime }})\exp \left(
\frac{ie}{c}\int\limits_{\vec{r}}^{\vec{r^{\prime
}}}\vec{A}d\vec{r}\right) ,
\end{equation}
where $G^{(0)}(i\epsilon _{n},\vec{r}-\vec{r^{\prime }})$ is the Green
function without magnetic field. In the zero order approximation in
the superconducting fluctuations, the disorder averaged Green function
$G^{(0)}$ in the momentum representation has the form:
\begin{equation}
G^{\left( 0\right) }\left( i\varepsilon _{n},{\bf p}\right) =\left(
i\varepsilon _{n}-\xi \left( {\bf p}\right) +i\left( 2\tau \right)
^{-1}sgn\varepsilon _{n}\right) ^{-1} \label{a5}
\end{equation}
The diagrams in Fig. 2 contain the averaged one-particle Green
functions, the impurity vertices proportional to the so called
Cooperon $C$ and the propagator of the superconducting fluctuations
$K$. The functions $C$ and $K$ depend on the coordinates and time
slower than the averaged one-particle Green functions because the
characteristic scale for both the impurity vertices and the propagator
of superconducting fluctuations $K$ is the coherence length $\xi $
which is much larger than $l$. As a result, the magnetic field affects
only the vertex $C$ and the propagator $K$, whereas the phases of the
Green functions drawn in Fig.2 outside these blocks cancel. So,
reading the diagrams in Fig.1 one should replace the solid lines by
the functions $G^{\left( 0\right) }\left( i\varepsilon _{n},{\bf p}
\right) $. More complicated diagrams containing crossings of impurity
lines describe the weak localization effects are neglected here.

The impurity vertex entering these diagrams is equal to $\left( 2\pi
\nu _{0}\tau \right) ^{-1}C\left( i\varepsilon _{n},i\Omega
_{k}-i\varepsilon _{n}\right) $, Fig. 3, where $\tau $ is the mean
free time due to the scattering on impurities or on the grain
boundary, $C$ is the Cooperon. It obeys the following equation
\begin{equation}
\left( D_{0}\left( -i{\bf \nabla -}\left( 2e/c\right) {\bf A}\right)
^{2}+|2\varepsilon _{n}-\Omega_{k}|\right) C\left( {\bf r,r}^{\prime
}\right) =2\pi \nu _{0}\delta \left( {\bf r-r}^{\prime }\right)
\label{a6}
\end{equation}
where $D_{0}=v_{0}^{2}\tau /3$ is the classical diffusion
coefficient. The vector-potential ${\bf A}\left( {\bf r}\right) $
should be chosen in the London gauge. If the shape of the grain is
close to spherical, the vector-potential is expressed through the
magnetic field ${\bf H}$ as ${\bf A}\left( {\bf r}\right) =[{\bf
H\times r]/}2$.
\begin{figure}
\epsfysize =2.5cm
\centerline{\epsfbox{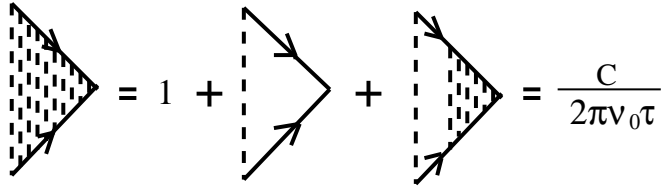}}
\caption{Impurity vertex.}
\end{figure}

All relevant energies in the problem are assumed to be much smaller
than the energy of the first harmonics $E_{c}=D_{0}\pi ^{2}/R^{2}$
playing the role of the Thouless energy of a single grain and this
allows to keep only the zero harmonics in the spectral expansion of
the solution $C\left( {\bf r,r}^{\prime }\right) $ of
Eq. (\ref{a6}). One can find the eigenvalue ${\cal E}_{0}\left(
H\right) $ of this harmonics using the first order of the standard
perturbation theory
\begin{equation}
{\cal E}_{0}\left( H\right) =\left( 2e/c\right) ^{2}D_{0}<{\bf
A}^{2}>_{0}
\label{a8}
\end{equation}
where $<...>_{0}$ stands for the averaging over the volume of the
grain. For the grain of a nearly spherical form one obtains
\begin{equation}
{\cal E}_{0}\left( H\right) =\frac{2}{5}\left( \frac{eHR}{c}\right)
^{2}D_{0}=\frac{2}{5}\left( \frac{\phi }{\pi \phi _{0}}\right)
^{2}E_{c}
\label{a9}
\end{equation}
where $\phi _{0}=\pi c/e$ is the flux quantum and $\phi $ is the
magnetic flux through the granule.

Within the zero-harmonics approximation, the function $C$ does not
depend on coordinates and equals
\begin{equation}
C\left( i\varepsilon _{n},i\Omega _{k}-i\varepsilon _{n}\right) =2\pi
\nu _{0}\left( |2\varepsilon _{n}-\Omega _{k}|+{\cal E}_{0}\left(
H\right) \right) ^{-1} \label{a10}
\end{equation}

To calculate the propagator of the superconducting fluctuations $K$
one should sum the sequence of the ladder diagrams represented in
Fig. 4. The broken lines in this figure denote the electron-electron
interaction. As it has been mentioned the characteristic energies of
the propagator $K$ are low and therefore, when calculating the
function $K$, one should take into account the tunneling processes
from grain to grains. The tunneling Hamiltonian $H_{T}$,
Eq. (\ref{tun}), is represented in Fig. 4 by crossed circles.
\begin{figure}
\epsfysize = 2.5cm
\centerline{\epsfbox{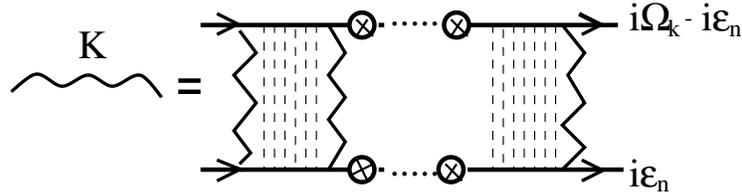}}
\caption{Propagator of superconducting fluctuations with tunneling.}
\label{fig4}
\end{figure}
Of course, one can sum the ladder diagrams in Fig.4 directly. However,
sometimes it is more convenient to decouple the electron-electron
interaction in Eq. (\ref{hamilton}) by a Gaussian integration over an
auxiliary field $\Delta $ (Hubbard-Stratonovich transformation). Then,
one can perform averaging over the electron quantum states, thus
reducing the calculation to computation of a functional integral over
the field $\Delta $. In principle, one obtains within such a scheme a
complicated free energy functional and the integral cannot be
calculated exactly. The situation simplifies if the fluctuations are
not very strong. Then, one can expand the free energy functional in
$\Delta $ and come to Gaussian integrals that can be treated without
difficulties. For the problem considered the propagator $K $ is
proportional to the average of the square of the field $\langle {
|\Delta _{k}|}^{2}\rangle $. In terms of the functional integral this
quantity is written as
\begin{equation}
\langle {|\Delta |}^{2}\rangle =\frac{\int {|\Delta |}^{2}\exp (-\beta
F_{eff}[\Delta ,\Delta ^{\ast }])D\Delta D\Delta ^{\ast }}{\int \exp
(-\beta F_{eff}[\Delta ,\Delta ^{\ast }])D\Delta D\Delta ^{\ast }}
\label{Delta}
\end{equation}
here $\beta =1/T$ and $F_{eff}[\Delta ,\Delta ^{\ast }]$ is the
effective free energy functional. We have chosen the parameters in
such a way that the grains are zero-dimensional. Therefore, it is
sufficient to integrate over the zero space harmonics only, which
means that the field $\Delta $ in the integral in Eq. (\ref{Delta})
does not depend on coordinates. In the quadratic approximation in the
field $\Delta $ the free energy functional $F_{eff}$ includes two
different terms
\begin{equation}
F_{eff}=F_{eff}^{(1)}+F_{eff}^{(2)}  \label{F1000}
\end{equation}
where $F_{eff}^{(1)}$ describes the superconducting fluctuations in an
isolated grain and $F_{eff}^{(2)}$ takes into account tunneling from
grain to grain. For the first term we obtain after standard
manipulations
\begin{equation}
F_{eff}^{(1)}\left[ \Delta ,\Delta ^{\ast }\right] =V\sum_{\Omega
_{k}}\left( 1/\left| \lambda \right| -T\sum_{2\varepsilon _{n}>\left|
\Omega _{k}\right| }2C\left( i\varepsilon _{n}\right) \right) {|\Delta
(\Omega _{k})|}^{2} \label{F11}
\end{equation}
where the function $C(i\varepsilon _{n})$ is defined in Eq.(\ref{a10})
and $V $ is the volume of a single grain. In the limit of low
temperatures $T\ll {\cal E}_{0}(H)$ the sum over the frequencies
$\varepsilon _{n}$ in Eq. (\ref {F11}) can be replaced by the integral
and we reduce the functional $F_{eff}^{(1)}\left[ \Delta ,\Delta
^{\ast }\right] $ to the form
\begin{equation}
F_{eff}^{(1)}\left[ \Delta ,\Delta ^{\ast }\right] =\frac{1}{\delta }
\sum_{\Omega _{k}}\left( \ln \left( \frac{{\cal E}_{0}(H)+|\Omega
_{k}|}{\Delta _{0}}\right) \right) {|\Delta (\Omega _{k})|}^{2}
\label{F1}
\end{equation}
Close to the critical magnetic field $H_{c}$ destroying the
superconducting gap in a single grain the energy of the first
harmonics ${\cal E}_{0}(H)$ is equal to the BSC gap at zero
temperature $\Delta _{0}$. This means that Eq. (\ref{F1}) can be
written in this case in the region $T\ll \Delta _{0}$. Near $H_{c}$
small frequencies $\Omega _{k}$ are most important and one can expand
Eq. (\ref{F1}) in powers of the small parameter $\Omega _{k}/\Delta
_{0}$. Then, we obtain
\begin{equation}
F_{eff}^{(1)}\left[ \Delta ,\Delta ^{\ast }\right] =\frac{1}{\delta }
\sum_{\Omega _{k}}\left( \ln \left( \frac{{\cal E}_{0}(H)}{\Delta
_{0}}\right) +\frac{|\Omega _{k}|}{\Delta _{0}}\right) {|\Delta
(\Omega _{k})|}^{2} \label{F12}
\end{equation}
At strong magnetic fields $H\gg H_{c}$, one should use the more
general formula, Eq. (\ref{F1}).

The term $F_{eff}^{(2)}\left[ \Delta ,\Delta ^{\ast }\right] $
describing the tunneling includes three different contributions
represented in Fig. 5
\begin{figure}
\epsfysize =4.5cm
\centerline{\epsfbox{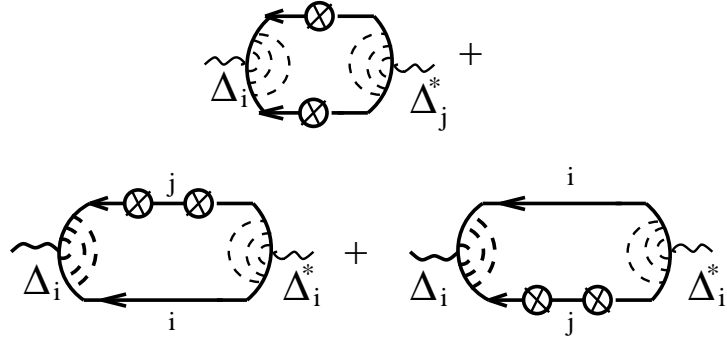}}
\caption{Diagrams describing $F_{eff}^{(2)}$.}
\label{fig5}
\end{figure}
The analytical expression $F_{eff}^{(21)}\left[ \Delta ,\Delta ^{\ast
}\right] $ corresponding to the first diagram Fig. 5 can be written
as
\begin{equation}
F_{eff}^{(21)}\left[ \Delta ,\Delta ^{\ast }\right] =-t^{2}/(2\pi \nu
_{0}\tau )^{2}\sum_{i,j}(\Delta _{i}\Delta _{j}^{\ast }+c.c.)V^{2}\int
\frac{d^{3}{\bf p}_{i}d^{3}{\bf p}_{j}}{{(2\pi
)}^{6}}T\sum_{\varepsilon _{n}}G(i\varepsilon _{n}{\bf
p}_{i})G(-i\varepsilon _{n}{\bf p} _{i})G(i\varepsilon _{n}{\bf
p}_{j})G(-i\varepsilon _{n}{\bf p} _{j})C^{2}(i\varepsilon _{n})
\label{F200}
\end{equation}
Writing Eq. (\ref{F200}) we put $\Omega _{k}=0$ in the expression for
the Cooperon $C$ and in the Green functions. This is justified because
the energy $F_{eff}^{\left( 2\right) }\left[ \Delta ,\Delta ^{\ast
}\right] $ is already small because it includes the parameter
$J(\delta /{\cal E}_{0}(H))$ that is assumed to be small, where $J$ is
the dimensionless conductance of the system specified by
Eq. (\ref{a40}). Next terms of the expansion are of the order of
$J(\delta /{\cal E}_{0}(H))(\Omega _{k}/{\cal E}_{0}(H))$ and can be
neglected for small $\Omega _{k}$. The second and third diagrams in
Fig. 5 are equal to each other and have the opposite sign with respect
to the first diagram. For simplicity we assume that the granules are
packed into a cubic lattice. Using the momentum representation with
respect to the coordinates of the grains and taking into account all
diagrams in Fig. 5 we reduce the free energy functional
$F_{eff}^{\left( 2\right) }\left[ \Delta ,\Delta ^{\ast }\right] $ to
the form
\begin{equation}
F_{eff}^{(2)}\left[ \Delta ,\Delta ^{\ast }\right] =(8/3\pi )(1/\delta
)\sum_{i=1}^{3}J(\delta /{\cal E}_{0}(H))\left( 1-\cos q_{i}d\right)
{|\Delta |}^{2} \label{F2}
\end{equation}
where ${\bf q}$ is the quasi-momentum and $d=2R$. Eq. (\ref{F2}) is
written in the limit
\begin{equation}
J\ll {\cal E}_{0}(H)/\delta  \label{F201}
\end{equation}

The inequality (\ref{F201}) is compatible with the inequality
(\ref{a400}) provided the inequality
\begin{equation}
{\cal E}_{0}(H)\gg \delta   \label{F202}
\end{equation}
is fulfilled. If ${\cal E}_{0}(H)\sim \Delta _{0}$, the condition,
Eq. (\ref {F202}), is at the same time the condition for the existence
of the superconducting gap in the single granule. The inequality
(\ref{F201}) allows us also to neglect influence of the tunneling on
the form of the Cooperon, so we use for calculations Eq. (\ref{a10}).

Writing the previous equations for $F_{eff}^{(2)}\left[ \Delta ,\Delta
^{\ast }\right] $ we neglected the influence of magnetic field on the
phase of the order parameter $\Delta $. In other words we omitted the
phase factor $\exp (i\frac{e}{c}\int {\bf A}({\bf r})d{\bf r})$. The
effect of the magnetic field on the phase will be discussed in details
later in Sec. \ref {sec6.5}.

Although the final result for the correction to the DOS can be written
for arbitrary temperatures $T$ and magnetic fields $H$, let us
concentrate on the most interesting case $T\ll T_{c}$,
$H>H_{c}$. Using Eqs. (\ref{Delta},\ref{F1},\ref{F2}) we obtain for
the propagator of the superconducting fluctuations $K\left( i\Omega
_{k,}{\bf q}\right) $

\[
K(i\Omega _{k},{\bf q})=-\nu _{0}^{-1}\left( \ln \left( \frac{{\cal E}
_{0}(H)+\left| \Omega _{k}\right| }{\Delta _{0}}\right) +\eta ({\bf q}
)\right) ^{-1},
\]

\begin{equation}
\eta ({\bf q})\equiv \left( 8/3\pi \right) \sum_{i=1}^{3}J\left(
\delta /{\cal E}_{0}(H)\right) \left( 1-\cos q_{i}d\right) ,
\label{eta}
\end{equation}
The pole of the propagator $K(i\Omega _{k},{\bf q})$ at ${\bf q}=0$,
$\Omega _{k}=0$ determines the field $H_{c}$, at which the BCS gap
disappears in a single grain. From the form of Eq. (\ref{eta}) we find
\begin{equation}
{\cal E}_{0}(H_{c})=\Delta _{0}.  \label{critical}
\end{equation}

The result for $H_{c}$, Eqs. (\ref{a9}, \ref{critical}), agrees with
the one obtained long ago by another method \cite{Larkin}. We can see
from Eqs. (\ref {eta}, \ref{critical}) that the term $\eta ({\bf q})$
describing tunneling is very important if $H$ is close to $H_{c}$.

Eqs. (\ref{a10}), (\ref{eta}) give the explicit formulae for the
functions $C $ and $K$ and allow us to calculate the correction
$\delta \nu $ to the DOS . The analytical expression for the diagram,
Fig. 2a, reads as follows
\begin{equation}
\delta \nu (i\varepsilon _{n})=(1/\pi )1/{(2\pi \nu _{0}\tau
)}^{2}\int \frac{d^{3}{\bf q}}{{(2\pi )}^{3}}T\sum_{\Omega
_{k}}K(i\Omega _{k},{\bf q})\int \frac{d^{3}{\bf p}}{{2\pi
}^{3}}C^{2}(i\varepsilon _{n},i\Omega _{k}-i\varepsilon
_{n})G^{2}(i\varepsilon _{n},{\bf p})G(i\Omega _{k}-i\varepsilon
_{n},{\bf p}) \label{diag1}
\end{equation}
Eq. (\ref{diag1}) contains integration over the momentum in the single
grain ${\bf p}$ and the quasi-momentum ${\bf q}$. First, we integrate
over the momentum ${\bf p}$ and reduce Eq. (\ref{diag1}) for
$\varepsilon _{n}>0$ to the form
\begin{equation}
\delta \nu (i\varepsilon _{n})=\frac{2iT}{\nu _{0}}\sum\limits_{\Omega
_{k}<\varepsilon _{n}}\int K(i\Omega _{k},{\bf q})C^{2}\left(
i\varepsilon _{n},i\Omega _{k}-i\varepsilon _{n}\right)
\frac{d^{3}{\bf q}}{\left( 2\pi \right) ^{5}} \label{fulnu}
\end{equation}
After calculation of the sum over $\Omega _{k}$ in Eq. (\ref{fulnu}),
one should make the analytical continuation $i\varepsilon
_{n}\rightarrow \varepsilon $. At low temperatures, it is sufficient
to find the correction to the DOS at zero energy $\delta \nu \equiv
\delta \nu \left( 0\right) $.

Remarkably, Eqs. (\ref{eta}-\ref{fulnu}) do not contain explicitly the
mean free time $\tau $. This is a consequence of the zero-harmonics
approximation, which is equivalent to using the random matrix theory
(RMT) Ref.~\cite{Efetov}. (The parameter $\tau $ enters only
Eq. (\ref{a9}) giving the standard combination ${\cal E}_{0}\left(
H\right) $ describing in RMT the crossover from the orthogonal to the
unitary ensemble). This justifies the claim that the results can be
used also for clean grains with a shape providing a chaotic electron
motion.

\begin{figure}
\epsfysize =3cm
\centerline{\epsfbox{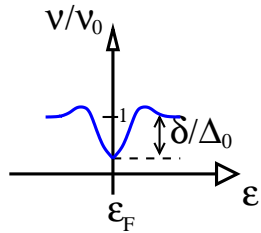}}
\caption{Suppression of DOS due to superconducting fluctuations.}
\label{fig6}
\end{figure}

Using Eqs. (\ref{a3}, \ref{eta}-\ref{fulnu}) one can easily obtain an
explicit expression for $\sigma _{DOS}$ for $H-H_{c}\ll H_{c}$. In
this limit, one expands the logarithm in the denominator of
Eq.(\ref{eta}) and neglects the dependence of $C$ on $\varepsilon
_{n}$ and $\Omega _{k}$ because the main contribution in the sum over
$\Omega _{k}$ comes from $\Omega _{k}\sim {\cal E}_{0}(H)-{\cal
E}_{0}(H_{c})\ll \Delta _{0}$. Using Eq. (\ref{connection}) the result
for $\delta \sigma _{DOS}=\sigma _{0}-\sigma _{DOS}$ can be written as
\begin{equation}
\frac{\delta \sigma _{DOS}}{\sigma _{0}}=-\frac{2\delta }{\Delta _{0}}
\left\{
\begin{array}{lr}
-\pi ^{-1}<\ln \tilde{\eta}\left( {\bf q}\right) >_{q}, & T/\Delta _{0}\ll 
\tilde{\eta} \\ 
\frac{2T}{\Delta _{0}}<\tilde{\eta}^{-1}({\bf q})>_{q}, & \tilde{\eta}\ll
T/\Delta _{0}\ll 1
\end{array}
\right. ,  \label{Hc}
\end{equation}
\[
\tilde{\eta}({\bf q})=\eta ({\bf q})+2h,\quad <...>_{q}\equiv
V\int\limits_{0}^{2\pi /d}\left( ...\right) d{\bf q/}\left( 2\pi
\right) ^{3}
\]
where $h=\left( H-H_{c}\right) /H_{c}$ and $V$ is a volume of the
single grain. We see that the correction to the conductivity is
negative and its absolute value decreases when the magnetic field
increases. The correction reaches its maximum at $H\rightarrow
H_{c}$. At zero temperature and close to the critical field $H_{c}$
such that $J(\delta /\Delta _{0})\gg h$, the maximum value of $\delta
\sigma _{DOS}/\sigma _{0}$ from Eq. (\ref{Hc}) is
\begin{equation}
\left| \frac{\delta \sigma _{DOS}}{\sigma _{0}}\right| =\frac{2}{\pi
}\frac{\delta }{\Delta _{0}}\left\langle \ln \left( \frac{1}{\eta
({\bf q})}\right) \right\rangle _{q}=\frac{1}{3}\frac{\delta }{\Delta
_{0}}\ln \left( \frac{\Delta _{0}}{J\delta }\right).
\label{estDOS0}
\end{equation}
In the limit $J(\delta /\Delta _{0})\ll h\lesssim 1$, one can expand
the logarithm in Eq. (\ref{Hc}). Then, taking $h\sim 1$ the
correction to the conductivity at zero temperature can be estimated as
\begin{equation}
\left| \frac{\delta \sigma _{DOS}}{\sigma _{0}}\right| \sim J\left(
\frac{\delta }{\Delta _{0}}\right) ^{2} \label{a22}
\end{equation}
Schematically, the suppression of the DOS due to the superconducting
fluctuations is shown in Fig. \ref{fig6}.
As temperature grows, the correction to the conductivity due to the
reduction of the DOS can become larger and reach for $T\sim \Delta
_{0}$ and $J(\delta /\Delta _{0})\gg h$ the order of magnitude of
$J^{-1}$.
\begin{equation}
\frac{|\delta \sigma _{DOS}|}{\sigma _{0}}=4(\delta T/\Delta
_{0}^{2})\left\langle \frac{1}{h+\eta ({\bf q})}\right\rangle _{q}\sim
\frac{1}{J} \label{estDOS}
\end{equation}
In the limit $h\approx 1\gg J(\delta /\Delta _{0})$ and at temperature
$T\sim \Delta _{0}$ this correction can be estimated as
\begin{equation}
\frac{\delta \sigma _{DOS}}{\sigma _{0}}\approx \frac{\delta }{\Delta
_{0}}
\label{a23}
\end{equation}

We see from Eqs. (\ref{estDOS0}-\ref{a23}) that the corrections to the
conductivity are smaller than unity provided we work in the regime of
a good metal, Eqs. (\ref{a1}, \ref{a400}), so the diagrammatic
expansion we use is justified. Indeed, we can neglect the corrections
of higher orders. For example, the diagram shown in Fig. 7 has the
additional small factor of $(\delta /\Delta _{0})\ln \tilde{\eta}$
at $T/\Delta _{0}\ll \tilde{\eta}$ and $\delta /(\Delta
_{0}\tilde{\eta})$ at $\tilde{\eta}\ll T/\Delta _{0}$.
\begin{figure}
\epsfysize =3cm
\centerline{\epsfbox{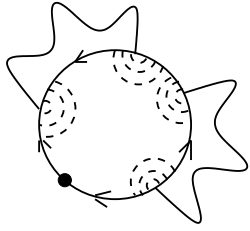}}
\caption{An example of high order corrections to DOS.}
\end{figure}
The correction to the conductivity calculated in this Section could
become comparable with $\sigma _{0}$ when $J\sim 1$. However, such
values of $J$ mean that we would be in this case not far from the
metal-insulator transition. Then, we would have to take into account
all localization effects. For values of $J\sim 1$ one can use
Eq. (\ref{Hc}) for rough estimates only. Apparently, the parameters of
the samples of Ref. \cite {Gerber97} correspond to the region $J\sim
1$, $\delta /\Delta _{0}\sim 1/3$.

In the limit of strong magnetic fields $H\gg H_{c}$ the correction to
$\sigma _{0}$ can still be noticeable. In this case we can use
Eq. (\ref{F1}) as before but, with a logarithmic accuracy, we can
neglect the dependence of the superconducting propagator,
Eq.(\ref{eta}), on $\Omega _{k}$ and on the tunneling term. Then we
obtain finally
\begin{equation}
\delta \sigma _{DOS}/\sigma _{0}=-(1/3)\left( \delta /{\cal
E}_{0}(H)\right) \ln ^{-1}\left( {\cal E}_{0}(H)/\Delta _{0}\right)
\label{highfield}
\end{equation}
Eq. (\ref{highfield}) shows that in the region $H\gg H_{c}$ the
correction to the conductivity decays essentially as $\delta \sigma
_{DOS}\sim H^{-2}$.

Let us emphasize that the correction to the conductivity coming from
the DOS remains finite in the limit $T\rightarrow 0$, thus indicating
the existence of the virtual Cooper pairs even at $T=0$.

In the region of not very small $h\gg J\delta /\Delta _{0}$, we can
neglect the tunneling term $F_{eff}^{\left( 2\right) }\left[ \Delta
,\Delta ^{\ast }\right] $ in the free energy functional
$F_{eff}\left[ \Delta ,\Delta ^{\ast }\right] $. Then, we can write
the correction to the conductivity in a rather general form. The
superconducting propagator $K$ can be written in this case as
\begin{equation}
K(i\Omega _{k})=-\nu _{0}^{-1}\ln ^{-1}\left( \frac{{\cal
E}_{0}(H)+|\Omega _{k}|}{\Delta _{0}}\right)
\end{equation}
Using Eq. (\ref{fulnu}) for the correction to the DOS and Eq. (\ref
{connection}) we obtain for the correction to the conductivity at zero
temperature
\begin{equation}
\frac{\delta \sigma _{DOS}}{\sigma _{0}}=-\frac{1}{3}\left(
\frac{\delta }{\Delta _{0}}\right) \int\limits_{a}^{\infty
}\frac{\exp {(-x)}}{x}dx=-\frac{1}{3}\left( \frac{\delta }{\Delta
_{0}}\right) Ei(a),\hspace{1cm}a=2\ln (1+h)
\end{equation}
In the limit $h\ll 1$, we reproduce with logarithmic accuracy
Eq. (\ref {estDOS0}), whereas in the opposite limiting case $h\gg 1$
we come to Eq. (\ref{highfield}).

In order to calculate the entire conductivity, Eq. (\ref{a2}), we must
investigate the AL and MT contributions (Figs. 2c and 2b). In
conventional superconductors near $T_{c}$, these contributions are
most important leading to an increase of the conductivity. In the
granular materials, the situation is much more interesting. It turns
out that both the AL and MT contributions {\it vanish} in the limit
$T\rightarrow 0$ at all $H>H_{c}$ and thus, the correction to the
conductivity comes from the DOS only. So, at low temperatures,
estimating the total correction to the classical conductivity $
\sigma _{0}$, Eq. (\ref{a4}), one can use the formulae of this
Section.

\section{Aslamasov-Larkin correction to the conductivity}

\label{sec3}

The Aslamasov-Larkin (AL) correction to the conductivity $\sigma
_{AL}$ originates from the ability of virtual Cooper to carry an
electrical current. In contrast to the one-electron tunneling
determining $\sigma _{DOS} $, the probability of tunneling of the
Cooper pairs from one grain to another is proportional to $t^{4}$. The
quantity $\sigma _{AL}$ is related to the response function $Q^{AL}$
as
\[
\sigma _{AL}=Q^{AL}/(-i\omega ),\omega \rightarrow 0 
\]
where the diagram for the $Q^{AL}(i\omega _{n})$ is represented in
Fig. 2c.  Calculating integrals corresponding to this diagram we may
put $\Omega _{k}=0 $ in the electron loops, because all singularities
in the vicinity of the transition point are contained in the
propagator of the superconducting fluctuations $K(i\Omega _{k})$
\cite{Aslamazov68}. The analytical expression for the diagram in
Fig. 2c has the form
\begin{equation}
Q^{AL}(i\omega _{n})=4/3\sum_{i=1}^{3}\int \frac{d^{3}{\bf q}}{{(2\pi
)}^{3}}T\sum_{\Omega _{k}}K(i\Omega _{k},{\bf q})K(i\Omega
_{k}-i\omega _{n},{\bf q})B_{1}^{2}(0,{\bf q}) \label{QAL}
\end{equation}
where $B_{1}(0,{\bf q})$ corresponds to one electron loop. The
analytical expression for this loop reads
\begin{eqnarray}
B_{1}(0,{\bf q})=- &&4i\int \sin q_{i}^{\prime }d\cos
(q_{i}-q_{i}^{\prime })d\frac{d^{3}{\bf q}^{\prime }}{{(2\pi
)}^{3}}\frac{edt^{2}V^{2}}{{(2\pi \nu _{0}\tau )}^{2}}\times
\label{c100} \\ &&\int \frac{d^{3}{\bf p}_{1}d^{3}{\bf p}_{2}}{{(2\pi
)}^{6}}T\sum_{\varepsilon _{n}}G(i\varepsilon _{n},{\bf
p}_{1})G(-i\varepsilon _{n},{\bf p}_{1})G(i\varepsilon _{n},{\bf
p}_{2})G(-i\varepsilon _{n},{\bf p}_{2})C^{2}(i\varepsilon _{n})
\nonumber
\end{eqnarray}
The functions $\sin q'_{i}d$ and $\cos (q_{i}-q'_{i})d$ in
Eq. (\ref{c100}) correspond to the current and tunneling vertices,
respectively. Eq. (\ref {QAL}) is obtained by considering four
different types of AL diagrams that are obtained from each other by
permutations of the current and tunneling vertices. Summing over the
spin of the electrons we get the additional factor $2$. As in the
preceding Section we calculate the impurity vertices neglecting the
tunneling term, which is justified if the inequality (\ref {F201}) is
fulfilled. Integrating over the momenta ${\bf p}_{1}$ and ${\bf p}
_{2}$ in Eq. (\ref{c100}) we reduce the functions $B_{1}(0,{\bf q})$
to the following form
\begin{equation}
B_{1}(0,{\bf q})=-\frac{8}{\pi ^{2}}J\frac{ed}{{\cal E}_{0}(H)}4i\int
\sin q_{i}d\cos (q_{i}-q_{i}^{\prime })d\frac{d^{3}{\bf q}^{\prime
}}{{(2\pi )}^{3}} \label{Bloop}
\end{equation}
To calculate the response function $Q^{AL}$ for real frequencies
$\omega $ one has to make an analytical continuation from the
Matsubara frequencies $\omega _{n}$. This can be done rewriting the
sum over $\Omega _{k}$ in Eq. (\ref{QAL}) in a form of a contour
integral that allows to make the continuation $i\omega _{n}\rightarrow
\omega +i0$. As a result, we obtain \cite{Aslamazov68}
\begin{eqnarray}
T\sum_{\Omega _{k}} && K\left( i\Omega _{k},{\bf q}\right)K\left(
i\Omega _{k}-i\omega _{n},{\bf q}\right) \rightarrow \label{sum} \\ &&
\frac{1}{4\pi i}\left( \int_{-\infty }^{+\infty }\coth
\frac{\varepsilon }{2T}(K^{R}(\varepsilon )-K^{A}(\varepsilon
))K^{A}(\varepsilon -\omega )d\varepsilon +\int_{-\infty }^{+\infty
}\coth \frac{\varepsilon }{2T}(K^{R}(\varepsilon )-K^{A}(\varepsilon
))K^{R}(\varepsilon +\omega )d\varepsilon \right)
\nonumber
\end{eqnarray}
where $K^{R}(K^{A})$ is retarded (advanced) superconducting
fluctuation propagator. Expanding Eq. (\ref{sum}) in $\omega $ we keep
the term that remains finite in the limit $\omega \rightarrow 0$ and
the linear one. The zero order term cancels with the contribution of a
diagram schematically represented in Fig.2c but containing instead of
	the current vertices the tunneling ones. The latter contribution
originates from the second term in Eq. (\ref{j}). The linear term
giving the dc conductivity can be written as
\begin{equation}
T\sum_{\Omega _{k}}K\left( i\Omega _{k},{\bf q}\right) K\left( i\Omega
_{k}-i\omega _{n},{\bf q}\right) \rightarrow -i\frac{2\omega {\cal E}
_{0}^{2}(H)}{\pi \nu _{0}^{2}}\int_{0}^{+\infty }\coth
\frac{\varepsilon }{2T}\frac{\varepsilon ({\varepsilon }^{2}-{\eta
}^{2})}{{({\eta }^{2}+{\varepsilon }^{2})}^{3}}d\varepsilon
\label{coth}
\end{equation}
where $\eta ({\bf q})$ is defined in Eq. (\ref{eta}). Using
Eqs. (\ref{QAL}, \ref{Bloop},\ref{coth}) the fluctuational
contribution $\sigma _{AL}$ to the conductivity is reduced in the
limit $T\ll T_{c}$, $H-H_{c}\ll H_{c}$ to the form
\begin{equation}
\frac{\sigma _{AL}}{\sigma _{0}}=\frac{16}{9}J\frac{\delta
^{2}T}{\Delta _{0}^{3}}\sum\limits_{i=1}^{3}<A\left( {\bf q}\right)
[{\tilde{\eta}({\bf q})]}^{-3}{\ \sin }^{2}q_{i}d>_{q} \label{fulAL}
\end{equation}
where $A\left( {\bf q}\right) =4\pi T\left( 3\Delta
_{0}\tilde{\eta}({\bf q})\right) ^{-1}$ for $T\ll \Delta
_{0}\tilde{\eta}$ and $A\left( {\bf q}\right) =1$ for $\Delta
_{0}\tilde{\eta}\ll T<T_{c}$.

Using Eq. (\ref{fulAL}) we can estimate the quantity $\sigma
_{AL}/\sigma _{0} $.

Let us consider first the limit $T\ll \Delta _{0}\tilde{\eta}$. In
this region, provided the inequality $h\ll J\delta /\Delta _{0}$ is
fulfilled, the main contribution in Eq. (\ref{fulAL}) comes from small
$q$. Calculating the integral over ${\bf q}$ we obtain from
Eq. (\ref{fulAL})
\begin{equation}
\frac{\sigma _{AL}}{\sigma _{0}}=\frac{32\pi ^{2}}{71}J\frac{\delta
^{2}T^{2}}{{\Delta _{0}}^{4}}\int_{0}^{\infty
}\frac{x^{2}d^{3}x}{\left( h+J(\delta /\Delta _{0})x^{2}\right)
^{4}}\sim J^{-3/2}\frac{T^{2}}{\Delta _{0}^{3/2}\delta ^{1/2}}\left(
\frac{H_{c}}{H-H_{c}}\right) ^{3/2}
\label{estALT0}
\end{equation}
From Eq. (\ref{estALT0}) we can see that at low temperatures the AL
correction to the conductivity is proportional to the square of the
temperature {\it \ }$\sigma _{AL}\sim T^{2}$ and vanishes in the limit
$T\rightarrow 0$.

Let us compare the AL correction with correction due to suppression of
the DOS considered in the previous Section. Using Eqs. (\ref{estALT0})
and (\ref {estDOS0}) we obtain
\begin{equation}
\frac{\sigma _{AL}}{|\delta \sigma _{DOS}|}\sim
J^{-3/2}\frac{T^{2}}{\Delta _{0}^{1/2}\delta ^{3/2}}\ln ^{-1}\left(
\frac{\Delta _{0}}{J\delta }\right) \left(
\frac{H_{c}}{H-H_{c}}\right) ^{3/2}.  \label{2}
\end{equation}
We see from Eq. (\ref{2}) that at $T\ll \Delta _{0}\tilde{\eta}$ the
AL correction is small $|\sigma _{AL}/\delta \sigma _{DOS}|\ll
1$. This means that the AL contribution cannot change the monotonous
increase of the resistivity of granulated superconductors when
decreasing the magnetic field. At very strong magnetic fields $H\gg
H_{c}$, we can neglect the second term in the denominator of
Eq. (\ref{estALT0}). Then the AL correction can be estimated as
\begin{equation}
\frac{\sigma _{AL}}{\sigma _{0}}\sim J\frac{\delta ^{2}T^{2}}{\Delta
_{0}^{4}}\left( \frac{H_{c}}{H}\right) ^{4}
\end{equation}
To compare this result with the correction due to the DOS we should
use Eq. (\ref{highfield}) that was also derived at strong magnetic
field $H\gg H_{c}$.
\begin{equation}
\left| \frac{\sigma _{AL}}{\delta \sigma _{DOS}}\right| \sim
J\frac{\delta T^{2}{\cal E}_{0}(H)}{\Delta _{0}^{4}}\left(
\frac{H_{c}}{H}\right) ^{4}\ln \left( \frac{{\cal E}_{0}(H)}{\Delta
_{0}}\right) \label{13}
\end{equation}
Using Eq. (\ref{a9}) for ${\cal E}_{0}(H)$ we can see from Eq. (\ref{13})
that $|\sigma _{AL}/\delta \sigma _{DOS}|_{H\gg H_{c}}\sim H^{-2}$.

Now let us consider the region of temperatures not far from the
critical temperature $\Delta _{0}\tilde{\eta}\ll T\leq T_{c}$. Using
Eq. (\ref{fulAL}) we have
\begin{equation}
\frac{\sigma _{AL}}{\sigma _{0}}=\frac{8\pi }{27}J\frac{\delta
^{2}T}{\Delta _{0}^{3}}\int \frac{Vd^{3}q\sin ^{2}qd}{\left(
h+J(\delta /\Delta _{0})(1-\cos qd)\right) ^{3}} \label{estAL}
\end{equation}
In the limit $J(\delta /\Delta _{0})\gg h$, the main contribution in
Eq. (\ref{estAL}) comes from small $q$ and the contribution $\sigma
_{AL}$ can be written as
\begin{equation}
\frac{\sigma _{AL}}{\sigma _{0}}=\frac{8\pi }{27}J(\delta ^{2}/{\Delta
_{0}}^{2})\int \frac{x^{2}d^{3}x}{\left( h+J(\delta /\Delta
_{0})x^{2}\right) ^{3}}\sim J^{-3/2}\left( \frac{\Delta _{0}}{\delta
}\right) ^{1/2}\left( \frac{H_{c}}{H-H_{c}}\right) ^{1/2}
\label{estAL2}
\end{equation}

We see that the AL contribution grows when approaching the critical
field $H_{c}$. In order to calculate the AL correction we used the
perturbation theory. Therefore, the region of the validity of the
results obtained is described by the inequalities $\sigma _{AL}/\sigma
_{0}\ll 1$ or $h\gg J^{-3}(\Delta _{0}/\delta )$.

Now let us compare the AL correction with $\delta \sigma _{DOS}$. From
Eqs. (\ref{estAL}, \ref{estDOS}) we obtain
\begin{equation}
\frac{\sigma _{AL}}{|\delta \sigma _{DOS}|}\sim J^{-1/2}\left(
\frac{\Delta _{0}}{\delta }\right) ^{1/2}\left(
\frac{H_{c}}{H-H_{c}}\right) ^{1/2}.
\label{1}
\end{equation}
Eq. (\ref{1}) is correct only at fields close to the critical field,
such that the inequality $h\ll J^{-1}(\Delta _{0}/\delta )$ is
fulfilled. In this region the total correction to the conductivity is
positive, which means the resistivity decays, when approaching the
critical magnetic field $H_{c}$. In the case of a strong magnetic
field $H\gg H_{c}$, we can neglect the second term in the denominator
of Eq. (\ref{estAL}) and then we obtain
\begin{equation}
\frac{\sigma _{AL}}{\sigma _{0}}\sim J\left( \frac{\delta }{\Delta
_{0}}\right) ^{2}\left( \frac{H_{c}}{H}\right) ^{3}
\end{equation}
that is $\sigma _{AL}\sim H^{-3}$ at $H\gg H_{c}$ and $T\sim \Delta
_{0}$.

To understand the behavior of the total conductivity of the granulated
superconductors in this region we should consider also the
Maki-Thompson contribution and this will be done in the next Section.

\section{Maki-Thompson correction to the conductivity}

\label{sec4}

Another contribution usually increasing the conductivity is the
Maki-Thompson (MT) contribution represented in Fig. 2b. Again, we can put $%
\Omega _{k}=0$ in the electron loop, because characteristic frequencies in
superconducting fluctuation propagator $K$ are of the order $\Omega _{k}\sim
\Delta _{0}\tilde{\eta}\ll {\cal E}_{0}$. The analytical expression for this
diagram reads 
\begin{equation}
Q^{MT}(i\omega _{n})=2/3\sum_{i=1}^{3}\int \frac{d^{3}{\bf q}}{{(2\pi
)}^{3}}T\sum_{\Omega _{k}}K(i\Omega _{k},{\bf q})B_{2}(i\omega
_{n},{\bf q})
\label{QMT}
\end{equation}
where $B_{2}(i\omega _{n},{\bf q})$ is a function describing the
contribution of the loop. This function can be written as follows
\begin{eqnarray}
&&B_{2}(i\omega _{n},{\bf q})=4\int \sin q_{i}^{\prime }d\sin
(q_{i}-q_{i}^{\prime })d\frac{d^{3}{\bf q}^{\prime }}{{(2\pi
)}^{3}}\frac{e^{2}d^{2}t^{2}V^{2}}{{(2\pi \nu _{0}\tau )}^{2}}\times
\nonumber \\ &&\int \frac{d^{3}{\bf p}_{1}d^{3}{\bf p}_{2}}{{(2\pi
)}^{6}}T\sum_{\varepsilon _{n}}G(-i\varepsilon _{n},{\bf
p}_{1})G(i\varepsilon _{n}, {\bf p}_{1})G(-i\varepsilon _{n}+i\omega
_{n},{\bf p}_{1})G(i\varepsilon _{n}-i\omega _{n},{\bf
p}_{2})C(i\varepsilon _{n})C(-i\varepsilon _{n}+i\omega
_{n},i\varepsilon _{n}-i\omega _{n}) \label{Bloop2}
\end{eqnarray}
where, as before, ${\bf p}_{1}$ and ${\bf p}_{2}$ stand for the
momenta in the granules and ${\bf q}$, ${\bf q}^{\prime }$ are
quasi-momenta. Eq. (\ref {QMT}) includes the contribution of two
different MT diagrams and summation over spins. In order to make the
analytical continuation $i\omega _{n}\rightarrow \omega +i0$ in
Eqs. (\ref{QMT}), (\ref{Bloop2}) it is convenient to rewrite the sum
over $\varepsilon _{n}$ in the form of the following contour integral
\begin{eqnarray}
T\sum_{\varepsilon _{n}}...=\frac{1}{4\pi i}( &\int_{C_{1}}&\tanh
\frac{z}{2T}G^{A}(-z,{\bf p}_{1})G^{R}(z,{\bf p}_{1})G^{A}(-z+i\omega
_{n},{\bf p}_{2})G^{R}(z-i\omega _{n},{\bf
p}_{2})C^{R}(z)C^{R}(z-i\omega _{n})dz+ \nonumber \\
&\int_{C_{2}}&\tanh \frac{z}{2T}G^{A}(-z,{\bf p}_{1})G^{R}(z,{\bf p}
_{1})G^{R}(-z+i\omega _{n},{\bf p_{2}})G^{A}(z-i\omega _{n},{\bf
p_{2}})C^{R}(z)C^{A}(z-i\omega _{n})dz+ \nonumber \\
&\int_{C_{3}}&\tanh \frac{z}{2T}G^{R}(-z,{\bf p}_{1})G^{A}(z,{\bf
p_{1}})G^{R}(-z+i\omega _{n},{\bf p}_{2})G^{A}(z-i\omega _{n},{\bf
p}_{2})C^{A}(z)C^{A}(z-i\omega _{n})dz), \label{countur}
\end{eqnarray}
where the contours $C_{1},C_{2},C_{3}$ are shown in Fig. 8.
\begin{figure}
\epsfysize =3.5cm
\centerline{\epsfbox{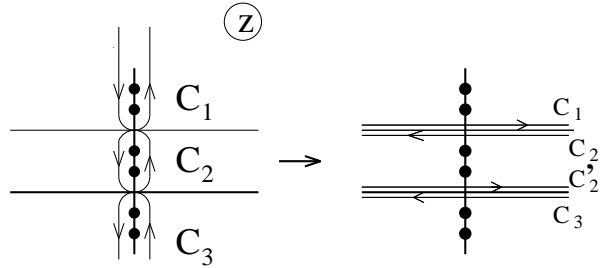}}
\caption{Contour of integration.}
\label{fig7}
\end{figure}
As usual, the MT diagrams have both regular (contours $C_{1},C_{3}$)
and anomalous (contour $C_{2}$) part. Near $T_{c}$ and at very low
magnetic field the anomalous part can be very large. In the limit
$H\rightarrow 0$, it can even diverge and become larger that the AL
correction giving a positive contribution to the conductivity
\cite{Varlamov}. However, in the limit of high magnetic fields and
$T\rightarrow 0$, the situation is less intriguing. It turns out that
for the problem considered, the absolute values of the regular and
anomalous parts are equal in this limit but these contributions have
the opposite signs. Making the analytical continuation in
Eq. (\ref{countur}) and integrating over the momenta we obtain in the
lowest order in $\omega $
\begin{equation}
\frac{1}{4\pi i}\left( \int_{C_{1}}+\int_{C_{2}}+\int_{C_{3}}\right)
\rightarrow \frac{8\pi i\omega }{{\delta }^{2}}\int_{0}^{+\infty
}\tanh \frac{\varepsilon }{4T}\frac{\varepsilon ({\cal
E}_{0}^{2}-{\varepsilon }^{2})}{({\cal E}_{0}^{2}{+{\varepsilon
}^{2}})^{3}}d\varepsilon \label{ccc}
\end{equation}
where the energy ${\cal E}_{0}={\cal E}_{0}\left( H\right) $ is given
by Eq.  (\ref{a9}). From Eq. (\ref{ccc}), we see that at $T=0$ the MT
contribution vanishes, which corresponds to the cancelation of the
regular and anomalous parts. At low but finite temperatures $T$, the
final result for the MT contribution can be written as
\begin{equation}
\frac{\sigma _{MT}}{\sigma _{0}}=\frac{16{\pi }^{2}T^{2}\delta
}{9\Delta _{0}^{3}}\sum\limits_{i=1}^{3}<B\left( {\bf q}\right) \cos
q_{i}d>_{q}
\label{fulMT}
\end{equation}
where $B\left( {\bf q}\right) =$ $-\pi ^{-1}\ln \tilde{\eta}({\bf q})$
for $T\ll \Delta _{0}\tilde{\eta}$ and $B\left( {\bf q}\right)
=2T\left( \Delta _{0}\tilde{\eta}({\bf q})\right) ^{-1}$ for $\Delta
_{0}\tilde{\eta}\ll T\lesssim T_{c}$

Let us estimate the MT contribution in the different limiting
cases. At low temperatures $T\ll \Delta _{0}\tilde{\eta}$ and $h\ll
J\delta /\Delta _{0}$ we have from Eq. (\ref{fulMT})
\begin{equation}
\frac{\sigma _{MT}}{\sigma _{0}}=\frac{16\pi }{9}\frac{T^{2}\delta
}{\Delta _{0}^{3}}\left\langle \ln \left( \frac{1}{h+J(\delta /\Delta
_{0})(1-\cos q)}\right) \cos q\right\rangle _{q}\sim \frac{T^{2}\delta
}{\Delta _{0}^{3}}
\label{estMT0}
\end{equation}
Comparing Eq. (\ref{estMT0}) with Eq. (\ref{estDOS0}) we come to the
following estimate
\begin{equation}
\frac{\sigma _{MT}}{|\delta \sigma _{DOS}|}\sim \left( \frac{T}{\Delta
_{0}}\right) ^{2}\ln ^{-1}\left( \frac{\Delta _{0}}{J\delta }\right)
\label{5.9}
\end{equation}
If the magnetic field $H$ is not very close to $H_{c}$, such that
$J\delta /\Delta _{0}\ll h\lesssim 1$, we obtain from
Eq. (\ref{fulMT})
\begin{equation}
\frac{\sigma _{MT}}{|\delta \sigma _{DOS}|}\sim \left( \frac{T}{\Delta
_{0}}\right) ^{2} \label{1000}
\end{equation}
We can see from Eqs. (\ref{5.9}, \ref{1000}) that the MT contribution
$\sigma _{MT}$ is proportional at low temperatures to $T^{2}$, which
is the same temperature dependence as that for the AL
contribution. This means that, at sufficiently low temperatures, the
MT contribution is small, $\sigma _{MT}/\delta \sigma _{DOS}\ll
1$. Thus, we conclude that, in this region, the main correction to the
classical conductivity, Eq. (\ref{a4}) comes from the correction to
the DOS, Eq. (\ref{estDOS0}). The latter correction is negative, so
the resistivity of the granulated superconductors exceeds its
classical value.

It is interesting to compare the AL and MT corrections. \ In the limit
$ T\rightarrow 0$, and $h\ll J\delta /\Delta _{0}$, we obtain using
Eqs. (\ref {estALT0}) and (\ref{estMT0})
\begin{equation}
\frac{\sigma _{MT}}{\sigma _{AL}}\approx \left( J\frac{\delta }{\Delta
_{0}}\right) ^{3/2}\left( \frac{H-H_{c}}{H_{c}}\right) ^{3/2}
\label{1001}
\end{equation}

Eq. (\ref{1001}) shows that in this region, The AL contribution is
larger the MT one. Let us consider another case of not very low
temperatures, $\Delta _{0}\tilde{\eta}\ll T\leq T_{c}$. From
Eq. (\ref{fulMT}) we have
\begin{equation}
\frac{\sigma _{MT}}{\sigma _{0}}=\frac{16\pi
^{3}}{27}\frac{T^{2}\delta }{\Delta _{0}^{3}}\int \frac{Td^{3}x\cos
x}{\Delta _{0}\left( h+J(\delta /\Delta _{0})(1-\cos x)\right) }\sim
\frac{1}{J} \label{estMT}
\end{equation}
Eq. (\ref{estMT}) gives the possibility to compare the MT contribution
with the DOS in this temperature interval. Recalling
Eq. (\ref{estDOS}) we obtain
\begin{equation}
\frac{\sigma _{MT}}{|\delta \sigma _{DOS}|}\sim 1  \label{1002}
\end{equation}
Eq. (\ref{1002}) shows that at not very low temperatures the MT
contribution has the same order of magnitude as the contribution due
to the reduction of the DOS. At the same time, the AL contribution in
the region $\Delta _{0}\tilde{\eta}\ll T\lesssim T_{c}$ can be
considerably larger than both the MT and DOS contributions.

From Eqs. (\ref{estAL2}) and (\ref{estMT}) we can see that at $T\lesssim
T_{c}$ 
\begin{equation}
\frac{\sigma _{MT}}{\sigma _{AL}}\approx J^{1/2}\left( \frac{\delta
}{\Delta _{0}}\right) ^{1/2}\left( \frac{H-H_{c}}{H_{c}}\right)
^{1/2}.  \label{1003}
\end{equation}
Eqs. (\ref{1002},\ref{1003}) show that, at not very low temperatures,
and not far from the critical field the AL correction to the
conductivity is the most important. This means that approaching the
transition in this region the resistivity decreases, which is in
contrast to the behavior at very low temperature where the correction
to the resistivity is determined entirely by the contribution to the
DOS and is positive.

To conclude the last two Sections we emphasize once more that the
temperature and magnetic field dependence of $\sigma _{AL}$ and
$\sigma _{MT} $ is rather complicated but they are definitely
positive. The competition between these corrections and $\sigma
_{DOS}$ determines the sign of the magnetoresistance. We see from
Eqs. (\ref{fulAL}, \ref{fulMT}) that both the AL and MT contributions
are proportional at low temperatures to $T^{2}$. Therefore the $\sigma
_{DOS}$ in this limit is larger and the magnetoresistance is negative
for all $H_{c}$. In contrast, at $T\sim T_{c}$ and close to $H_{c}$,
the AL and MT corrections can become than $\sigma _{DOS}$ resulting in
a positive magnetoresistance in this region. Far from $H_{c}$ the
magnetoresistance is negative again.

\section{The critical field $H_{c_{2}}$ in the granulated superconductors}

\label{sec6.5}

In the previous Sections we considered transport in granulated
superconductors at magnetic fields $H$ not far from the field
$H_{c}$. The field $H_{c}$ is the field destroying the superconducting
gap in a single isolated granule. We have seen that the main
contribution due to the superconducting fluctuations comes from the
correction to the density of states, Eq. (\ref{Hc}), and this
correction remains finite in the limit $H\rightarrow H_{c}$. But is
the field $H_{c}$ a critical field in the system of the granules
coupled to each other by tunneling? If it were a critical field what
would happen at $H<H_{c}$? Would the system be macroscopically the
superconductor or normal metal? Or, may be, this would be a new state
of matter?

To answer these questions we should derive the effective action
$F_{eff}\left[ \Delta ,\Delta ^{\ast }\right] $, Eqs. (\ref{F1000},
\ref{F1}, \ref {F2}), more carefully than it has been done in
Sec.\ref{sec2}. Namely, until now we considered only the effect of the
magnetic field on the electron motion inside the grains, neglecting
its influence on the correlation of the phases of the order parameter
$\Delta _{i}$ of different grains. However, to understand whether the
system is macroscopically superconductor or not, we must consider the
macroscopic motion and thus, the effect of the magnetic field on the
phase correlation.

In this Section we come back to the derivation of effective action
$F_{eff}\left[ \Delta ,\Delta ^{\ast }\right] $ taking into account
the influence of magnetic field on the phases of the order
parameter. First, we consider this problem qualitatively and then,
quantitatively. It is clear that if the magnetic field is strong
enough, it induces macroscopic currents that finally, at a field
$H_{c_{2}}$, destroy the superconductivity.

Let us estimate the critical magnetic field $H_{c_{2}}$ using the
Ginzburg-Landau theory. We assume that the granulated system under
consideration is in the macroscopically superconducting state. Then,
the Josephson part of the free energy in the coordinate
representation, not too close to the $H_{c_{2}}$, can be written in
the form
\begin{equation}
F=\sum\limits_{i}E_{J}\frac{(\Delta _{i}-\Delta _{j})^{2}}{\Delta
_{0}^{2}}\approx \int d^{3}{\bf r}\frac{(\nabla \Delta
)^{2}}{R\Delta _{0}^{2}}E_{J}
\label{c2}
\end{equation}
where $E_{J}=J\Delta _{0}$ is the Josephson energy, $J\gg 1$ is the
dimensionless conductance, Eq. (\ref{a40}), $\Delta _{0}$ is the BCS
gap at zero magnetic field and $R$ is a radius of the single
grain. The gradient expansion in Eq. (\ref{c2}) was done under the
assumption that the magnetic flux through one grain is smaller than
the flux quantum $\phi _{0}$.

In the conventional Ginzburg-Landau free energy the coefficient in
front of the gradient term is proportional to the square of the
coherence length $\xi $. Therefore, using Eq. (\ref{c2}) we can
extract the macroscopic coherence length $\xi $. Recalling that the
free energy of a single grain is $\nu V\Delta ^{2}$, where $V$ is a
volume of single grain, we obtain
\begin{equation}
\xi ^{2}\sim JR^{2}(\delta /\Delta _{0})  \label{c3}
\end{equation}
If the conductance $J$ is large enough the behavior of the granulated
superconductors is the same as in a bulk sample with the effective
coherence length $\xi $. Now we estimate the critical magnetic field
$H_{c_{2}}$ that destroys the superconductivity in the system as
$H_{c_{2}}\xi ^{2}\approx \phi _{0}$. This field is different from the
field $H_{c}$. Using Eq. (\ref {critical}) we can compare these two
fields. The result for the ratio of these two fields is
\begin{equation}
\frac{H_{c}}{H_{c_{2}}}\approx J\frac{\delta }{\Delta
_{0}}\sqrt{\frac{R}{\xi _{0}}} \label{c4}
\end{equation}
Eq. (\ref{c4}) for $H_{c_{2}}$ is valid provided $H_{c_{2}}\ll H_{c}$,
which corresponds to the inequality $J\gg (\Delta _{0}/\delta
)\sqrt{\xi _{0}/R}$.  However, we are interested in the opposite case
when
\begin{equation}
1\ll J\ll (\Delta _{0}/\delta )\sqrt{\xi _{0}/R}  \label{c5}
\end{equation}
This contradict to the assumption made and means that, in the region
specified by Eq. (\ref{c5}), the critical field $H_{c_{2}}$ is close
to the field $H_{c}$ and $\left| H_{c}-H_{c_{2}}\right| /H_{c}$ can be
considered as a small parameter.

Now let us calculate the critical magnetic field $H_{c_{2}}$ more
rigorously taking into account the influence of the magnetic field on
the macroscopic motion. The magnetic field results in an additional
phase factor $\exp (i\frac{e}{c}\int {\bf A(r)}d{\bf r})$ in the
superconducting field $\Delta _{i}$ in Eq. (\ref{F200}). We assume
that this magnetic field is not far from $H_{c}$ determined by
Eqs. (\ref{a9}, \ref{critical}). The field $H_{c}$ is of order $\phi
_{0}/R\xi $, which means that, in the limit under consideration $R\ll
\xi $, the magnetic flux through the elementary cell of the lattice of
the granules is small. Therefore, we can expand the functions $\exp
(i\frac{e}{c}\int {\bf A(r)}d{\bf r})$ in ${\bf A}$ and write
gradients instead of the finite differences of $\Delta _{i}$ in
Eq. (\ref {F200}). Essential frequencies $\Omega _{k}$ are also small
and the free energy functional in such a continuum approximation takes
the form
\begin{equation}
F\left[ \Delta ,\Delta ^{\ast }\right] =\frac{1}{\delta}
\sum\limits_{\Omega _{k}}\left( \ln \left( \frac{{\cal
E}_{0}(H)}{\Delta _{0}}\right) +\frac{\mid \Omega _{k}\mid }{{\cal
E}_{0}(H)}+\frac{4d^{2}}{3\pi }J\left( \frac{\delta }{{\cal
E}_{0}(H)}\right) \left( \nabla -\frac{2ie}{c}{\bf A}\right)
^{2}\right) \left| \Delta \right| ^{2} \label{c6}
\end{equation}
The critical magnetic field $H_{c_{2}}$ can be found writing the
propagator of the superconducting fluctuation $K$ corresponding to the
free energy, Eq.  (\ref{c6}). Making Fourier transformation of the
function $\Delta $ in the eigenfunctions of the operator entering
Eq. (\ref{c6}) and calculating Gaussian integrals we obtain for the
propagator $K_{n}\left( 0,q_{z}\right) $ in the spectral
representation at $\Omega _{k}=0$
\begin{equation}
K_{n}(0,q_{z})=-\nu _{0}^{-1}\left( \ln \left( \frac{{\cal
E}_{0}(H)}{\Delta _{0}}\right) +\frac{4d^{2}}{3\pi }J\left(
\frac{\delta }{{\cal E}_{0}(H)}\right) \left(
q_{z}^{2}+4(n+1/2)\frac{H}{\phi _{0}}\right) \right) ^{-1}
\label{c7}
\end{equation}
where $q_{z}$ is a component of the quasi-momentum parallel to the
magnetic field and $n=0,1...$ are the number of the Landau levels. To
calculate the critical magnetic field $H_{c_{2}}$\ we should consider
poles of the superconducting propagator. Taking the lowest Landau
number $n=0$ and putting $q_{z}=0$ obtain the following equation
determining the critical field $H_{c_{2}}$
\begin{equation}
\ln \left( \frac{{\cal E}_{0}(H_{c_{2}})}{\Delta _{0}}\right)
+\frac{8d^{2}}{3\pi }J\left( \frac{\delta }{\Delta _{0}}\right)
\frac{H_{c}}{\phi _{0}}=0
\label{correction}
\end{equation}
Expanding the first term in Eq. (\ref{correction}) near $H_{c}$ we
find the critical field $H_{c_{2}}$
\begin{equation}
H_{c_{2}}=H_{c}\left( 1-\frac{40}{\pi ^{2}}J\left( \frac{\delta
}{\Delta _{0}}\right) \sqrt{\frac{R^{2}}{\xi _{0}l}}\right)
\label{secondfield}
\end{equation}
Eq. (\ref{secondfield}) shows us that the critical field $H_{c_{2}}$
is close to the field $H_{c}$ so long as $\left( J\delta /\Delta
_{0}\right) \left( R/\xi \right) \ll 1,$ where $\xi \sim \left( \xi
_{0}l\right) ^{1/2}$ is the coherence length in the superconducting
grains. Eq. (\ref{F201}) and the assumption that the grains are
zero-dimensional guarantee the fulfillment of this inequality. For the
ballistic motion of \ electrons inside the grain (the radius of the
grain $R$ is of the order of the mean free path, $R\sim l$) the
critical field $H_{c_{2}}$ can be written in the form
\begin{equation}
\frac{H_{c}-H_{c_{2}}}{H_{c}}=\frac{40}{\pi ^{2}}J\left( \frac{\delta
}{\Delta _{0}}\right) \sqrt{\frac{R}{\xi _{0}}}\ll 1
\end{equation}
Eq. (\ref{secondfield}) is the main result of this Section. Below the
magnetic field $H_{c_{2}}$, one should add in Eq. (\ref{c6}) a term
quartic in $\Delta $, which gives a non-zero order parameter $\Delta
.$ Thus, at field $H<H_{c_{2}}$ the granular system is in the
superconducting state.

In order to understand the behavior of the resistivity as a function
of the magnetic field $H$ in the region $H_{c_{2}}<H<H_{c}$ we can
consider the quantity $\frac{\partial }{\partial H}\left( \frac{\delta
\sigma }{\sigma _{0}}\right) $. We can use as before Eq. (\ref{fulnu})
but now, calculating the integral, we should make the following
replacement
\[
\int \left( ...\right) \frac{d^{3}{\bf q}}{\left( 2\pi \right) ^{3}}
\rightarrow \frac{H}{\phi _{0}}\int \sum\limits_{n=0}^{\infty
}(...)\frac{dq_{z}}{2\pi }
\]
where $q_{z}$ is a component of the quasi-momentum parallel to the
magnetic field.

The correction to the DOS at $T=0$ takes the form 
\begin{equation}
\delta \nu (0)=\frac{1}{2\pi ^{2}\nu _{0}}C^{2}T\sum\limits_{\Omega
_{k}<0}\left( \frac{H}{\phi _{0}}\right) \sum\limits_{n=0}^{\infty
}\int K_{n}(i\Omega _{k},q_{z})\frac{dq_{z}}{(2\pi )} \label{c8}
\end{equation}
The main contribution to the correction $\delta \nu (0)$ in
Eq. (\ref{c8}) comes from the term with $n=0$. Using the fact that
$\delta \sigma /\sigma _{0}=2\delta \nu /\nu _{0},$ taking the first
derivative with respect to the magnetic field and finally, integrating
over the frequency $\Omega $ and quasi-momentum $q_{z}$ we obtain
\begin{equation}
\frac{\partial }{\partial H}\left( \frac{\delta \sigma }{\sigma
_{0}}\right) =-15\sqrt{\frac{\pi }{24}}\frac{1}{H_{c}}\left(
\frac{\delta }{J\Delta _{0}}\right) ^{1/2}\left( \frac{R^{2}}{\xi
_{0}l}\right) ^{1/2}\left( \frac{H_{c_{2}}}{H-H_{c_{2}}}\right)
^{1/2}, \label{derivative}
\end{equation}
where $H_{c_{2}}$ is given by Eq. (\ref{secondfield}). We can see from
Eq. (\ref{derivative}) that the value $\frac{\partial }{\partial
H}\left( \frac{\delta \sigma }{\sigma _{0}}\right) $ diverges when
$H$ approaches $H_{c_{2}} $. Thus, the critical field $H_{c_{2}}$ is
characterized by the infinite slope on the dependence of the
resistivity on the magnetic field.  This property might help to
identify this field on experimental curves.

\section{Zeeman splitting.}

\label{6.6}

In our previous consideration we neglected interaction between the
magnetic field and spins of the electrons. This approximation is
justified if the size of the grains is not very small. Then, the
critical field $H_{c}^{or \text{ }}$destroying the superconducting
gap is smaller than the paramagnetic limit $g\mu H_{c}^{Z}=\Delta
_{0}$ and the orbital mechanism dominates the magnetic field effect on
the superconductivity. However, the Zeeman splitting leading to the
destruction of the superconducting pairs can become important if one
further decreases the size of the grains.

Let us discuss now the effect of Zeeman splitting. We can rewrite
Eq. (\ref {orbit}) for the ratio of orbital $H_{c}^{or}$ magnetic
field to the Zeeman magnetic field in the following form
\begin{equation}
\frac{H_{c}^{or}}{H_{c}^{Z}}=\left(\frac{\delta }{\delta _{c}}\right)
^{1/3}
\label{twofield}
\end{equation}
where $\delta _{c}\approx 1/(\nu R_{c}^{3})$, $R_{c}=\xi
(p_{0}l)^{-1}$ and $\xi =\sqrt{\xi _{0}l}$. To understand whether
the Zeeman splitting is important for an experiment we can estimate
the ratio $\delta _{c}/\Delta _{0}$ and compare it with the proper
experimental result. We find easily
\begin{equation}
\frac{\delta _{c}}{\Delta _{0}}=(p_{0}l)\sqrt{\frac{l}{\xi _{0}}}
\label{delta_c}
\end{equation}
which shows $\delta _{c}/\Delta _{0}$ can be both smaller than
$(\delta /\Delta _{0})_{exper.}$ and larger depending on the values of
$l$ and $\xi _{0}$. Using the result for $\delta _{c}$,
Eq. (\ref{delta_c}), we can rewrite Eq. (\ref{twofield}) as
\begin{equation}
\frac{H_{c}^{or}}{H_{c}^{Z}}=\left( \frac{\delta }{\Delta
_{0}}\frac{R}{l}\right) ^{1/2}
\end{equation}

In the experiment \cite{Gerber97}, both the mechanisms are in
principle important. One can come to this conclusion using the fact
that the Zeeman critical magnetic field is $H_{c}^{Z}\approx 3.5T$ and
this is not far from the peak in the resistivity at the field
$H\approx 2.5T$. Below, we consider the corrections to the DOS and
conductivity due to the Zeeman mechanism. We will see that at
temperature $T=0$ these corrections can be of the same order of
magnitude as the correction due to orbital mechanism.

Let us calculate first the critical magnetic field $H_{0}$ destroying
the superconducting gap in a single grain taking into account both the
orbital and Zeeman mechanisms of the destruction. The Green function
for the non-interacting electrons in this case is
\begin{equation}
G_{\uparrow \downarrow }^{(0)}(i\varepsilon_n, {\bf
p})=\frac{1}{i\varepsilon _{n}-\xi ({\bf p})\pm E_{Z}/2+(i/2\tau
)sgn(\varepsilon _{n})}, \label{GF}
\end{equation}
where $E_{Z}=g\mu _{B}H$ is the Zeeman energy. Including the
interaction between the magnetic field and electron spins we obtain
the following form of the Cooperon
\begin{equation}
C(i\varepsilon _{n},-i\varepsilon _{n})=2\pi \nu _{0}(\left|
2\varepsilon _{n}\right| -iE_{Z}sgn\left( \varepsilon _{n}\right)
+{\cal E}_{0}(H))^{-1}
\label{cooperon2}
\end{equation}
Repeating the calculations of Sec. \ref{sec2} with the modified
Cooperon, Eq. (\ref{cooperon2}), we find at $T\ll Tc$ $\ $the new
critical magnetic field $H_{0}$
\begin{equation}
{\cal E}_{0}^{2}(H_{0})+E_{Z}^{2}(H_{0})=\Delta _{0}^{2}
\label{newfield}
\end{equation}
In the limit of a very week Zeeman splitting, when the orbital
mechanism is more important, Eq. (\ref{newfield}) reproduces the
previous result for the critical magnetic field Eq. (\ref{critical}). In
the opposite limiting case, when the Zeeman mechanism plays the major
role, we obtain $E_{Z}=\Delta _{0}$. In general case, when the both
mechanisms of the destruction of the conductivity are important, one
should solve Eq. (\ref{newfield}) and the result reads
\begin{equation}
H_{0}=H_{c}\left( -\frac{1}{2}\left( \frac{g\mu _{B}H_{c}}{\Delta _{0}}
\right) ^{2}+\sqrt{\frac{1}{4}\left( \frac{g\mu _{B}H_{c}}{\Delta
_{0}}\right) ^{4}+1}\right) ^{1/2} \label{general}
\end{equation}
Using Eq. (\ref{critical}) for a critical field $H_{c}$ we can
estimate the ratio $\mu _{B}H_{c}/\Delta _{0}=(\xi
_{0}/R)(p_{0}R)^{-1}(p_{0}l)^{-1}\ll 1$. If this parameter is small
(the orbital mechanism is more important) the critical field $H_{0}$
is close to the field $H_{c}$
\begin{equation}
H_{0}=H_{c}\left( 1-\frac{1}{4}\left( \frac{g\mu _{B}H_{c}}{\Delta
_{0}}\right) ^{2}\right)
\end{equation}
When Zeeman splitting is more important then orbital mechanism then
from Eq.  (\ref{general}) we obtain $g\mu _{B}H_{0}=\Delta _{0}$. This
is the point of the absolute instability of the paramagnetic state. At
$g\mu _{B}H<\Delta _{0}$, the superconducting state is the only stable
one.

Let us calculate the correction to the DOS taking into account both
the Zeeman splitting and orbital mechanism of the suppression of
superconductivity. Using Eqs. (\ref{GF}, \ref{cooperon2}) we obtain
for the superconducting propagator $K$
\begin{equation}
K(i\Omega _{k})=-\nu _{0}^{-1}\left( \frac{1}{2}\ln \left(
\frac{({\cal E}_{0}(H)+|\Omega _{k}|)^{2}+E_{Z}^{2}}{\Delta
_{0}^{2}}\right) +\eta ({\bf q})\right) ^{-1} \label{Kbig}
\end{equation}
In the region $\tilde{h}=(H-H_{0})/H_{0}\ll 1$, where $H_{0}$ is given
Eq. (\ref{general}), expanding the logarithm in the superconducting
propagator we obtain
\begin{equation}
K(i\Omega _{k})=-\nu _{0}^{-1}\left( 2\tilde{h}+\frac{|\Omega
_{k}|}{\Delta _{0}}+\eta ({\bf q})\right) ^{-1}
\end{equation}
Using Eqs. (\ref{fulnu}, \ref{connection}) for the correction to the
conductivity we obtain the same result as before, Eq. (\ref{Hc}), but
with the new $h_{0}$. Eqs. (\ref{estDOS0}-\ref{a23}) are correct for
in general case.

In the limit of a strong magnetic field $\tilde{h}\gg 1$ we can
neglect with logarithmic accuracy the $\Omega $-dependence of
superconducting propagator in Eq. (\ref{Kbig}). Using
Eqs. (\ref{fulnu}, \ref{cooperon2}, \ref {connection}) we obtain for
the correction to the conductivity at strong magnetic fields
\begin{equation}
\frac{\delta \sigma _{DOS}}{\sigma _{0}}=-\frac{2}{3}\left(
\frac{\delta }{E_{Z}(H)}\right) \arctan \left( \frac{E_{Z}(H)}{{\cal
E}_{0}(H)}\right) \ln ^{-1}\left( \frac{{\cal
E}_{0}^{2}(H)+E_{Z}^{2}(H)}{\Delta _{0}^{2}}\right)
\label{highfield2}
\end{equation}
From Eq. (\ref{highfield2}) we can see that if orbital mechanism is
more important than the Zeeman one, that is if $E_{Z}/{\cal E}_{0}\ll
1$, we reproduce the previous result for the correction to the
conductivity at strong magnetic field, Eq. (\ref{highfield}). If
Zeeman mechanism is more important $E_{Z}/{\cal E}_{0}\gg 1$, then we
obtain for the correction to the conductivity
\begin{equation}
\frac{\delta \sigma _{DOS}}{\sigma _{0}}=-\frac{\pi }{6}\left(
\frac{\delta }{E_{Z}(H)}\right) \ln ^{-1}\left( \frac{E_{Z}(H)}{\Delta
_{0}}\right)
\label{l}
\end{equation}
Eq. (\ref{1}) has the same structure as Eq. (\ref{highfield}) but the
function ${\cal E}_{0}(H)$ is replaced by $E_{Z}(H)$. This changes the
asymptotic behavior at strong magnetic fields because $E_{Z}\left(
H\right) \sim H$ in contrast to ${\cal E}_{0}\left( H\right) $ $\sim
H^{2}$. So, we conclude from Eq. (\ref{l}) that $\delta \sigma
_{DOS}/\sigma _{0}\sim H^{-1} $.

\section{Diamagnetic susceptibility of granular superconductors}

\label{sec6}

In previous sections we have demonstrated that the resistivity of the
granulated superconductors grows when approaching the superconducting
state from the region of very strong magnetic fields. Resistivity is a
quantity studied experimentally most often. Another quantity
accessible experimentally is the magnetic susceptibility. Can one
observe anything unusual measuring the dependence of the
susceptibility as a function of the magnetic field?

The diamagnetic susceptibility of a bulk sample above the critical
temperature $T_{c}$ in a weak magnetic field has been studied long ago
\cite {Schmid}. In this Section, we want to present results for the
diamagnetic susceptibility $\chi $ of the granular superconductors in
the opposite limit of strong magnetic fields and low temperatures. As
will be shown below, the fluctuations of virtual Cooper pairs always
increase the absolute value of the diamagnetic susceptibility $\chi $
of the granulated system. In contrast to the conductivity, the
diamagnetic susceptibility is determined mainly by currents inside the
granules and is finite even in isolated granules.  Therefore, the fact
that the virtual Cooper pairs cannot move from grain to grain, which
is crucial for the conductivity, is not very important for the
magnetic susceptibility and the latter does not show a non-monotonic
behavior characteristic for the resistivity.

To derive explicit formulae, let us consider first the limit of very
low temperature $T\ll T_{c}$ and strong magnetic field $H-H_{c}\ll
H_{c}$. The effective free energy functional $F\left[ \Delta ,\Delta
^{\ast },H\right] $ in the quadratic approximation in the order
parameter $\Delta $ has been already obtained for this case and is
given by Eqs. (\ref{F1000}, \ref{F12}, \ref{F2}). The diamagnetic
susceptibility $\chi $ can be calculated using the standard relations
\begin{equation}
\chi =-(1/V)\langle \partial ^{2}F/\partial H^{2}\rangle _{q}  \label{dia1}
\end{equation}
where the free energy $F$ has the form 
\begin{equation}
F=-T\ln \left( \int d\Delta d\Delta ^{\ast }\exp [-\beta
F_{eff}(\Delta ,\Delta ^{\ast },H)]\right) \label{dia2}
\end{equation}
where, as before, $V$ is the volume of a single grain and the
averaging $\langle ...\rangle _{q}$ is specified in
Eq. (\ref{Hc}). Calculating the derivative in Eq. (\ref{dia1}) and
using Eqs. (\ref{F1000}, \ref{F12}, \ref {F2}, \ref{dia2}) we find for
the diamagnetic susceptibility
\begin{equation}
\chi =-\left\langle \frac{1}{V}T\sum\limits_{\Omega _{k}}\left(
H-H_{c}+\frac{H_{c}}{2}\frac{\mid \Omega _{k}\mid }{\Delta
_{0}}+\frac{8}{3\pi }\frac{H_{c}}{2}J\left( \frac{\delta }{\Delta
_{0}}\right) \sum\limits_{i=1}^{3}(1-\cos q_{i}d)\right)
^{-2}\right\rangle _{q}
\label{sum2}
\end{equation}
At very low temperature $T/\Delta _{0}\ll h=\left( H-H_{c}\right)
/H_{c}$, we can replace the sum over frequency $\Omega _{k}$ by an
integral. As a result we obtain
\begin{equation}
\chi =-\frac{3}{5}\chi _{L}\left( \frac{l}{R}\right) \left\langle
\left( h+\frac{4}{3\pi }J\left( \frac{\delta }{\Delta _{0}}\right)
\sum\limits_{i=1}^{3}(1-\cos q_{i}d)\right) ^{-1}\right\rangle _{q}
\label{sum3}
\end{equation}
where $\chi _{L}=e^{2}v_{0}/12\pi ^{2}\hbar c^{2}$ is the Landau
diamagnetic susceptibility. In the limit $h\gg J(\delta /\Delta _{0})$
tunneling between grains is not important and the diamagnetic
susceptibility of the granulated system is equal to the susceptibility
of a single grain. If the motion of electrons inside a grain is more
or less ballistic ($l\approx R$) the result takes the form:
\begin{equation}
\chi (H)=-\frac{\pi }{10}\chi _{L}\left( \frac{H_{c}}{H-H_{c}}\right)
\approx -10^{-7}\left( \frac{H_{c}}{H-H_{c}}\right) \label{chi}
\end{equation}
From Eq. (\ref{chi}), we can see that the fluctuation-induced
diamagnetic susceptibility $\chi $ can appreciably exceed the value
$\chi _{L\text{ }}$ due to the Landau diamagnetism. In order to probe
the diamagnetism due to the superconducting fluctuations
experimentally one can measure the field-dependent part of
susceptibility and compare it with Eq. (\ref{chi}).

In the limit $h\ll T/\Delta _{0}$, it is sufficient to take into
account only one term with $\Omega_{k}=0$ in the sum in
Eq. (\ref{sum2}). Then, the result for the diamagnetic susceptibility
$\chi $ can be written in the form:
\begin{equation}
\chi =-\frac{\pi ^{2}}{5}\chi _{L}\left( \frac{l}{R}\right) \left(
\frac{T}{\Delta _{0}}\right) \left\langle \left( h+\frac{4}{3\pi
}J\left( \frac{\delta }{\Delta _{0}}\right)
\sum\limits_{i=1}^{3}(1-\cos q_{i}d)\right) ^{-2}\right\rangle _{q}
\label{chi1}
\end{equation}
In the limit when $h\gg J(\delta /\Delta _{0})$ and $l\approx R$
$,$Eq. (\ref {chi1}) can be simplified and one comes to the following
expression
\begin{equation}
\chi (H)=-\frac{\pi ^{3}}{30}\chi _{L}\frac{T}{\Delta _{0}}\left(
\frac{H_{c}}{H-H_{c}}\right) ^{2} \label{chi2}
\end{equation}
Eqs. (\ref{chi1}, \ref{chi2}) show that the diamagnetic susceptibility
$\chi $ diverges in a power law when magnetic field $H$ is close to
the field $H_{c}$ but the powers are different for the two different
regions of the fields.

At zero temperature $T=0$, as we can see directly from
Eq. (\ref{sum3}), the susceptibility $\chi $ remains finite even in
the limit $H\rightarrow H_{c}$. However, as we have discussed in
Sec. \ref{sec6.5}, the field $H_{c}$ does not correspond to any phase
transition. The transition to the superconductivity occurs at a lower
field $H_{c_{2}}$. So, it is interesting to consider the critical
behavior near $H_{c_{2}}$.

Proper calculations of the diamagnetic susceptibility in the region $
H_{c_{2}}<H<H_{c}$ and at temperature $T\ll \Delta _{0}$ can be
carried out without any difficulty. The free energy functional is
given in this case by Eq. (\ref{c6}) and we obtain for the diamagnetic
susceptibility $\chi $
\begin{equation}
\chi =-T\sum\limits_{\Omega _{k}}\frac{H_{c}}{\phi _{0}}\int
\frac{dq_{z}}{2\pi }\left( H-H_{c_{2}}+\frac{H_{c}}{2}\frac{|\Omega
_{k}|}{\Delta _{0}}+H_{c}\frac{2d^{2}}{3\pi }\frac{J\delta }{\Delta
_{0}}q_{z}^{2}\right)^{-2} \label{dia3}
\end{equation}
In the limit $T\rightarrow 0$ we replace the summation over $\Omega
_{k}$ by integration. Integrating over the frequency and the
quasi-momentum $q_{z}$ we obtain
\begin{equation}
\chi =-\frac{\pi ^{3/2}}{\sqrt{20}}\chi _{L}\left( \frac{\Delta
_{0}}{J\delta }\right) ^{1/2}\left( \frac{l}{\xi _{0}}\right)
^{1/2}\left( \frac{H_{c_{2}}}{H-H_{c_{2}}}\right) ^{1/2}
\label{dia4}
\end{equation}
We see from Eq. (\ref{dia4}) that the diamagnetic susceptibility $\chi
$ diverges in a power law when $H$ $\rightarrow H_{c_{2}}$ but the
power $1/2$ is different from those in Eqs. (\ref{chi}, \ref{chi2})
describing the behavior of the conductivity in the region $H>H_{c}$.

The case of temperatures close to the critical temperature $T-T_{c}\ll
T_{c}$ and weak magnetic fields $H\ll H_{c}$ has been considered for
bulk samples long ago \cite{Schmid}. So, we present here the result
for the diamagnetic susceptibility $\chi $ of the granulated
superconductors only in the limit $% J(\delta /T_{c})\gg
(T-T_{c})/T_{c}$ and $l\approx R$
\begin{equation}
\chi \approx -\chi _{L}\frac{T_{c}}{J\delta } \label{dia5}
\end{equation}
Eq. (\ref{dia5}) shows that the diamagnetic susceptibility of the
granular superconductors near $T_{c}$ is still larger than the
magnitude of the Landau diamagnetism.

In all previous considerations we did not take into account the spin
paramagnetism. When the size of the grains is large, $R\gg R_{c}\equiv
\xi (p_{0}l)^{-1}$, this effect are small in comparison with the
diamagnetism.

\section{Correction to conductivity at $\left| T-T_{c}\right| \ll T_{c}$}

\label{sec5}

Effect of superconducting fluctuations on DOS of isotropic bulk
samples has been considered in the limit $T-T_{c}\ll T_{c}$ and $H\ll
H_{c}$ long ago \cite{Abrahams}. The AL and MT contributions to the
conductivity were considered in the same limit for layered
superconductors in Ref.~\cite {Varlamov}. Here we want to extend these
results to the case of the granulated superconductors. Experimentally,
only a small increase of the resistivity $\delta \rho /\rho _{0}\sim
0.03-0.04$ has been observed in this region \cite{Gerber97}. The
reason of the reduction of the effect in the vicinity of the critical
temperature $T_{c}$ is that the AL and MT corrections are not small in
comparison with the correction from the DOS.  Repeating the
calculations of Sections \ref{sec2} and \ref{sec3} for $T-T_{c}\ll
T_{c}$ and $H=0$ we write the contribution from the DOS and the AL
correction as
\begin{equation}
\frac{\delta \sigma _{DOS}}{\sigma _{0}}=-\frac{7\zeta (3)}{2{\pi
}^{2}}\frac{\delta }{T_{c}}\left\langle
\frac{1}{\frac{T-T_{c}}{T_{c}}+\eta _{2}(q)}\right\rangle _{q},
\label{Tc}
\end{equation}

\begin{equation}
\frac{\sigma _{AL}}{\sigma _{0}}=0.17\frac{t^{2}}{T_{c}^{2}}
\sum\limits_{i=1}^{3}\left\langle \frac{\sin ^{2}q_{i}d}{{\ }\left(
{\frac{T-T_{c}}{T_{c}}+\eta _{2}(q)}\right) ^{3}}\right\rangle _{q}
\label{main}
\end{equation}
where $\eta _{2}(q)\equiv 1/3\sum\limits_{i=1}^{3}J(\delta
/T_{c})(1-\cos q_{i}d)$.

To calculate the MT correction, Fig. 2b, we should renormalize the
impurity vertices taking into account the tunneling term. This is
necessary, because in the anomalous MT contribution strongly diverges
in low dimension if the magnetic field is weak. So, the tunnelling
from grain to grain can provide convergence of the integrals giving
the anomalous MT contribution. The Dyson equation for the Cooperon $C$
can be written as
\begin{equation}
C=C_{0}+C_{0}\Sigma C  \label{cop}
\end{equation}
where all diagrams for the self-energy $\Sigma $ are shown in
Fig. 9. The function $C_{0}$ in Eq. (\ref{cop}) is the Cooperon in a
single grain.
\begin{figure}
\epsfysize =2.3cm
\centerline{\epsfbox{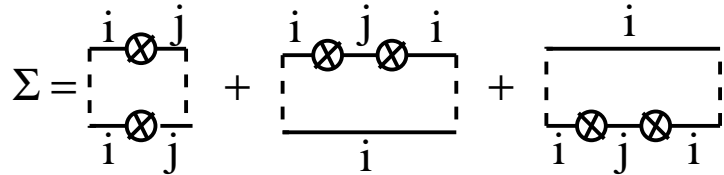}}
\caption{Diagrams contributing to the self-energy.}
\end{figure}

Solving Eq. (\ref{cop}) we see that the proper propagator for the
anomalous part of the MT correction has an additional diffusion pole
in comparison with the regular part. Therefore, in the limit $T_{c}\gg
$ $\eta _{2}(q)$, the anomalous MT contribution to the conductivity
$\sigma _{MT}^{(an)}$ is larger than the regular one $\sigma
_{MT}^{(reg)}$ and their ratio is proportional to $T_{c}/\eta
_{2}(q)$. At the same time, the anomalous contribution $\sigma
_{MT}^{(an)}$ is positive, which means that both the AL and MT
corrections give a positive contribution to the conductivity.
Explicit formulae for the regular and anomalous parts of the
conductivity can be written as

\begin{equation}
\frac{\sigma _{MT}^{(reg)}}{\sigma _{0}}=-\frac{7\zeta (3)}{48{\pi
}^{2}}\frac{\delta }{T_{c}}\sum\limits_{i=1}^{3}\left\langle
\frac{\cos q_{i}d}{(\frac{T-T_{c}}{T_{c}}+\eta
_{2}(q))}\right\rangle _{q} \label{regular}
\end{equation}

\begin{equation}
\frac{\sigma _{MT}^{(an)}}{\sigma _{0}}=0,06\frac{\delta }{T_{c}}
\sum\limits_{i=1}^{3}\left\langle \frac{\cos q_{i}d}{\eta
_{2}(q)(\frac{T-T_{c}}{T_{c}}+\eta _{2}(q))}\right\rangle _{q}
\label{TcMT}
\end{equation}
Eqs. (\ref{a2}, \ref{Tc}, \ref{main}, \ref{regular}, \ref{TcMT})
describe completely the behavior of the conductivity near $T_{c}$. We
see that the terms $\sigma _{AL}$ and $\sigma _{MT}^{(an)}$ giving
positive contributions to the conductivity diverge in the limit
$T\rightarrow T_{c}$, whereas the terms $\sigma _{DOS}$ and $\sigma
_{MT}^{\left( reg\right) }$ reducing the conductivity converge in this
limit. Therefore, sufficiently close to $T_{c}$, the superconducting
fluctuations increase the conductivity. A weak magnetic field shifts
the critical temperature $T_{c}$ and one can describe also the
dependence of the conductivity on the magnetic field. Apparently, far
from the transition point one can obtain an increase of the
resistivity due to the superconducting fluctuations and thus, a peak
in the resistivity.  However, this peak should be small, which
correlates with the experimental observation near
$T_{c}$\cite{Gerber97}. It is only the region of low temperatures
considered in the previous sections where a considerable negative
magnetoresistance is possible.

\section{Experiments on $Al$ grains.}

\label{sec7}

The theoretical study presented in this paper was motivated by the
experimental work \cite{Gerber97}. Let us compare the available
experimental results with our theory. In the article \cite{Gerber97}
three samples were studied. We concentrate our attention on the
samples 1 and 2, Fig. 4, of that work.

We analyze the case of very low temperatures $T\ll T_{c}$ and magnetic
fields $H>H_{c}$, where $T_{c}\approx 1.6K$ is the critical
temperature for $Al$ grains studied in the experiment and $H_{c}$ is
the critical magnetic field that suppresses the superconductivity in a
single grain, Eqs. (\ref{a9}, \ref{critical}). At temperature
$T\simeq 0.3K$ and magnetic field $H\simeq 4T$ these samples show a
large negative magnetoresistance. The resistivity of the sample 2 has
the maximum at $H=2.5T$ and the value of this peak is more than twice
as large as the resistivity in the normal state (that is, at $H\gg
H_{c}$, when all superconducting fluctuations are completely
suppressed). A negative magnetoresistance due to weak localization
(WL) effects is also not unusual in disordered metals and, to describe
the experimental data, its value should be estimated as well as the
effects of the superconducting fluctuations discussed in the previous
chapters.

The total conductivity of the granular metal under consideration
including effects of WL and superconducting fluctuations can be
written in the form:
\begin{equation}
\sigma =\sigma _{0}+\delta \sigma _{DOS}+\sigma _{AL}+\sigma
_{MT}+\delta \sigma _{WL} \label{totalcond}
\end{equation}
At low temperatures $T\ll T_{c}$, the contribution $\delta \sigma
_{DOS}$ originating from the reduction of DOS due to the formation of
the virtual Cooper pairs is larger than the contributions $\sigma
_{AL}$ and $\sigma _{MT}$ since the latter vanish in the limit
$T\rightarrow 0$. So, let us concentrate on estimating the
contributions $\delta \sigma _{DOS}$ and $\delta \sigma _{WL}$.

It is clear that the sample $1$ undergoes a metal-insulator
transition, which results in the complete suppression of the
superconductivity. The parameters of the sample 2 that is of the main
interest for us are not far from the those of the sample $1$. Using
Eq. (\ref{a40}), the value $R=60$\AA\ for the radius of the grains,
and the value of the resistivity $\rho _{0}\approx 1.9\times
10^{-2}\Omega cm$ we find $J\approx 0.1$. 

The small value of $J$ is not in the contradiction with the
possibility for the system to be in the metallic phase. The Anderson
metal-insulator transition in granular metals was considered using an
effective medium approximation \cite {efetov88,Efetov}. The critical
point $J_{c}$ in the present notations is given for the $3D$ cubic
lattice by the equation (Eq. (12.67) of the book \cite{Efetov})
\begin{equation}
\left( \frac{4J_{c}}{\pi }\right) ^{1/2}\ln
\left(\frac{1}{4J_{c}}\right)=\frac{1}{5}
\label{exp0}
\end{equation}
We can see from Eq. (\ref{exp0}) that the critical value of
$J_c=10^{-3}$ is really very small. Therefor, we believe that the
metal-insulator transition observed in the sample $1$ is not a
conventional Anderson transition. Apparantly it occurs due to
formation of the superconducting gap. Then, the transition can be
described following the scenario of Ref.\cite{efetov80}. 

Why can one be sure that the experimentally the weak localization
corrections are small?  A similar effect of the negative
magnetoresistance has been observed in Ref. \cite{Rutgers} and the
authors of that work attributed it to the weak localization effects.
Could it be that this effect is really due to the weak localization
corrections and the present theory is not relevant to the experiment?

However, it is not difficult to show that in the case under
consideration the weak localization corrections originating from a
contribution of Cooperons are totally suppressed by the magnetic
field. This is not in contradiction with the fact that the system is
close to the metal-insulator transition because strong localization is
possible even if the Cooperons are absent.

To calculate the contribution $\sigma _{WL}$ coming from the Cooperons
we extended the standard derivation of the correction \cite{cor} to the
case of the granulated metal. Using approximations developed
previously one can obtain without difficulties the following
expression for a 3-dimensional cubic lattice of the grains
\begin{equation}
\delta \sigma _{WL}=-\frac{16}{3}e^{2}J\delta
d^{2}\sum_{i=1}^{3}\int_{0}^{2\pi /d}\frac{C_{{\bf q}}\left( 0\right)
}{2\pi \nu _{0}}\cos q_{i}d\frac{d^{3}{\bf q}}{\left( 2\pi \right)
^{3}}
\label{exp1}
\end{equation}
where the function $C_{{\bf q}}\left( 0\right) $ is the Cooperon taken
at the frequency $\omega =0$ and quasi-momentum $q.$

What remains to do is to take the explicit expression for the Cooperon
and compute the integral over the quasi-momentum $q$. However, if we
use Eq. (\ref{a10}) derived previously we get identically zero. This
is because Eq. (\ref{a10}) was derived without taking into account
tunneling. We can modify Eq. (\ref{a10}) writing as in Sec. \ref{sec5}
the Dyson equation, Eq. (\ref {cop}), with $\Sigma $ represented in
Fig. 9. As a result, we obtain
\begin{equation}
C_{{\bf q}}\left( i\omega _{n}\right) =2\pi \nu _{0}\left( \left|
\omega _{n}\right| +\frac{16}{\pi }J\delta \sum_{i=1}^{3}\left( 1-\cos
q_{i}d\right) +{\cal E}_{0}\left( H\right) \right) ^{-1} \label{exp2}
\end{equation}
In the limit under consideration, Eq. (\ref{F201}), the second term in
the brackets in Eq. (\ref{exp2}) is much smaller that the third one
and one can expand the function $C_{{\bf q}}\left( i\omega _{n}\right)
$, considering the second term as a perturbation. Restricting
ourselves by the first order, substituting the result into
Eq. (\ref{exp1}) and using Eq. (\ref{a4}) we write the final result
for the correction $\delta \sigma _{WL}$ in the form
\begin{equation}
\frac{\delta \sigma _{WL}}{\sigma _{0}}=-\frac{8J}{\pi }\left(
\frac{\delta }{{\cal E}_{0}\left( H\right) }\right) ^{2} \label{exp3}
\end{equation}
Eq. (\ref{exp3}) shows that the weak localization correction in the
strong magnetic fields considered here is always small. In contrast,
the correction to the conductivity coming from the DOS
\begin{equation}
\frac{\delta \sigma _{DOS}}{\sigma _{0}}=-\frac{1}{3}\frac{\delta
}{\Delta _{0}}\ln \left( \frac{\Delta _{0}}{J\delta }\right)
\label{exp4}
\end{equation}
obtained in Sec. \ref{sec2} for $T=0$ can be considerably larger. The
ratio of these two corrections takes at $H\sim H_{c}$ the form
\begin{equation}
\frac{\delta \sigma _{WL}}{\delta \sigma _{DOS}}=\frac{24}{\pi }\left(
\frac{J\delta }{\Delta _{0}}\right) \ln ^{-1}\left( \frac{\Delta
_{0}}{J\delta }\right) \label{exp5}
\end{equation}
which must be small in the limit $J\ll \Delta _{0}/\delta $.

Now, let us estimate the corrections $\delta \sigma _{DOS}$ and
$\delta \sigma _{WL}$ using the parameters of the experiment
\cite{Gerber97}. For the typical diameter $120\pm 20\AA $ of $Al$
grains studied in \cite {Gerber97} the mean level spacing $\delta $ is
approximately $\delta \approx 1K$. Using the critical temperature
$T_{c}\simeq 1.6K$ for $Al$ we obtain for the BCS gap in a single
grain the following result $\Delta _{0}\approx 1.8T_{c}\approx
3K$. Substituting the extracted values of the parameters into
Eq. (\ref{exp4}) we can estimate the maximal increase of the
resistivity. As a result, we obtain $\left( \delta \rho /\rho
_{0}\right) _{\max }\approx 0.4$, which is somewhat smaller but not
far from the value $\left( \delta \rho /\rho \right) _{\exp }\approx
1$ observed experimentally.

Although our theory gives smaller values of $\left( \delta \rho /\rho
_{0}\right) _{\max }$ than the experimental ones, the discrepancy
cannot be attributed to the weak localization effects. Using the
experimental values of $J$, $\delta $ and $\Delta _{0}$ we find from
Eq. (\ref{exp5}) that $\delta \sigma _{WL}$ is $10$ times smaller
than $\delta \sigma _{DOS}$. The value of the correction $\delta
\sigma _{WL}/\sigma _{0}$, Eq. (\ref{exp3}), near $H_{c}$ equals
$2.8\times 10^{-2}$. Strictly speaking, all calculations have been
done under the assumption of a large $J\gg 1$, while experimentally
this parameter is not large. This means that our theory does not take
into account all effects that might be relevant for the experiment
\cite{Gerber97}. Possibly, at such small values of $J$ as one has in
the experiment charging effects become important reducing additionally
the density of states. However, study of the effects of the Coulomb
interaction is beyond the scope of the present paper.

We see from Eq. (\ref{highfield}) that, if the orbital mechanism of
the destruction of the superconductivity is more important than the
Zeeman one, then in the region of ultra high magnetic fields $H\gg
H_{c}$ the correction to the resistivity decays as $\delta \rho /\rho
_{0}\sim H^{-2}$. In the opposite limiting case, when the Zeeman
splitting is more important, we see from Eq. (\ref{l}) that $\delta
\rho /\rho _{0}\sim H^{-1}$. This means that the correction to the
classical conductivity is still sensitive to the magnetic field far
away from $H_{c}$. As concerns the weak localization correction
$\delta \sigma _{WL}$ given by Eq. (\ref{exp3}), it decays at large
magnetic fields as $H^{-4}$ and is always small.

The dependence of the conductivity $\sigma $ on the magnetic field $H$
is completely different at temperatures $\left| T-T_{c}\right| \ll
T_{c}$ because in this region all types of the corrections considered
in the previous sections can play a role depending on the value of the
magnetic field. At low magnetic fields the Aslamazov-Larkin $\delta
\sigma _{AL}$ and anomalous Maki-Thompson $\delta \sigma _{MT}^{\left(
an\right) }$ corrections give the main contribution because they are
most divergent near the critical point. At the same time, as the
magnetic field grows they decay faster than the correction $\delta
\sigma _{DOS}$ originating from the reduction of the density of
states. At a certain magnetic field $H^{\ast }$ all the corrections
can become of the same order of magnitude. In this region of the
fields, one can expect a negative magnetoresistance, although it
cannot be large. Comparing all the types of the corrections with each
other we estimate the characteristic magnetic field as $H^{\ast }\sim
H_{c}J^{-1}\left( \Delta _{0}/\delta \right) $. The theory presented
here was developed under the assumption $J^{-1}(\Delta _{0}/\delta
)\gg 1$ and, hence, the characteristic field is very large, $H^{\ast
}\gg H_{c}$.

Experimentally, the peak in the resistivity at temperatures close to
$T_{c}$ can be estimated as $\delta \rho /\rho _{0}\sim 3.5$ and is
much smaller than the peak at low temperatures. This correlates, at
least qualitatively, with our results.

\section{Conclusion}

In this paper we presented a detailed theory of the new mechanism of
the negative magnetoresistance in granular superconductors in a strong
magnetic field suggested recently \cite{Igor} to explain the
experiment \cite{Gerber97}. We considered the limit of a large
conductance, thus neglecting localization effects. It has been
demonstrated that even if the superconducting gap in each granule is
destroyed by the magnetic field the virtual Cooper pairs can persist
up to extremely strong magnetic fields.  However, the contribution of
the Cooper pairs to transport is proportional at low temperatures to
$T^{2}$ and vanishes in the limit $T\rightarrow 0$.  In contrast, they
reduce the one-particle density of states in the grains even at $T=0$,
thus diminishing the macroscopic conductivity. The conductivity can
reach its classical value only in extremely strong magnetic fields
when all the virtual Cooper pairs do not exist anymore. This leads to
the negative magnetoresistance.

We analyze both the orbital and Zeeman mechanisms of the destruction
of superconductivity as well as the limits of low temperatures and
temperatures close to the critical temperature $T_{c}$. The results
demonstrate that, at low temperatures $T\ll T_{c}$, there must be a
pronounced peak in the dependence of the resistivity on the magnetic
field. This peak should be much smaller in the region of temperatures
$T\ll T_{c}$ because, in this region the superconducting fluctuations
can contribute to transport, thus diminishing the role of the
reduction of the density of states. We were able also to consider
different regions of the magnetic fields.

\begin{figure}
\epsfysize =4.5cm
\centerline{\epsfbox{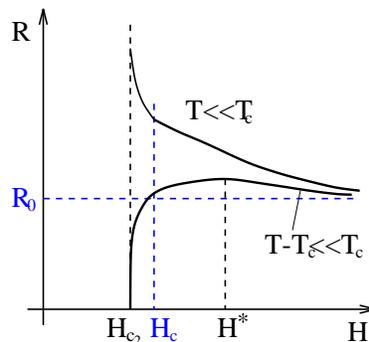}}
\caption{Resistivity of the granulated superconductors as a function
of the magnetic field at fixed temperatures in different regimes.}
\label{fig9}
\end{figure}

Qualitatively, the results are summarized in Fig. 9, where typical
curves for the low temperatures $T\ll T_{c}$ and temperatures close to
$T_{c} $ are represented. Both the functions reach asymptotically the
value of the classical resistivity $R_{0}$ only at extremely strong
magnetic fields. The resistivity $R$ at low temperatures grows
monotonously when decreasing the magnetic field. The function does not
have any singularity at the magnetic field $H_{c}$ destroying the
superconducting gap $\Delta $ in a single grain. The real transition
into the superconducting state occurs at a lower field
$H_{c_{2}}$. This field, in the region of parameters involved, is
close to the field $H_{c}$. The resistivity $R\left( H\right) $
remains finite as $H\rightarrow H_{c_{2}}$ but its derivative diverges
resulting in the infinite slope at $H=H_{c_{2}}$. The dependence of
the resistivity $R$ on the magnetic field at temperatures near $T_{c}$
is more complicated.  Already far from the field $H_{c}$ the
superconducting fluctuations start contributing to transport and the
resistivity goes down when decreasing the magnetic field. A negative
magnetoresistance is possible in this region only at magnetic fields
$H^{\ast }$, that can be much larger than the field $H_{c} $.

The theory developed gives a good description of existing experiments.
Although the experimental systems are close to the metal-insulator
transition and localization effects as well as Coulomb interaction can
play an essential role, our theory, where all these effects were
neglected, gives reasonable values of physical quantities and allows
to reproduce the main features of experimental curves.

It was important for our calculations that the dimensionless
conductance $J\gg 1$ of the sample was limited also from above, such
that the inequality $J\ll \Delta _{0}/\delta $ was fulfilled. This
means that the granulated structure of the superconductor was
essential for us. However, the fact that the contribution of the
superconducting fluctuations vanishes in the limit $T\rightarrow 0$
(AL and MT corrections are proportional to $T^{2}$) seems to be rather
general and not restricted by this inequality. Apparently, the
negative magnetoresistance above the critical magnetic field persists
and can be possible even in conventional bulk superconductors. We
leave the region of large conductances $J\gtrsim \Delta _{0}/\delta $
for a future study.

\section{ACKNOWLEDGMENTS}

The authors thank I. Aleiner, B. Altshuler, F. Hekking for helpful
discussions in the course of the work. A support of the
Graduiertenkolleg 384 and the Sonderforschungsbereich 237 is greatly
appreciated. The work of one of the authors (A.I. L.) was supported by
the NSF grant DMR-9812340. He thanks also the Alexander von Humboldt
Foundation for a support of his work in Bochum.

%\end{multicols} 

\end{document}